\documentclass[iop]{emulateapj}
\usepackage{graphicx}
\usepackage{amssymb}
\usepackage{threeparttable}

\mathchardef\mhyphen="2D

\newcommand{\oiii}{[O {\sc iii}]}

\newcommand{\feii}{Fe {\sc ii}}

\mathchardef\mhyphen="2D

\slugcomment{Accepted for publication in ApJS on September 16, 2015}

\shorttitle{Bridging IFU \& absorption line analyses}
\shortauthors{Liu et al.}

\begin{document}


\title{Integral Field Spectroscopy of AGN Absorption Outflows:\\ Mrk 509 and IRAS F04250$-$5718}


\author{
Guilin Liu\altaffilmark{1,$\dagger$}, 
Nahum Arav\altaffilmark{1}
and
David S. N. Rupke\altaffilmark{2}
}

\affil{$^1$Department of Physics, Virginia Tech, Blacksburg, VA 24061, USA}
\affil{$^2$Department of Physics, Rhodes College, Memphis, TN 38112, USA}

\altaffiltext{$\dagger$}{Email: glliu@vt.edu}


\begin{abstract}

Ultravoilet (UV) absorption lines provide abundant spectroscopic information 
enabling the probe of the physical conditions in AGN outflows, but the 
outflow radii (and the energetics consequently) 
can only be determined indirectly. We present the first direct test of these 
determinations using integral field unit (IFU) spectroscopy. We have 
conducted Gemini IFU mapping of the ionized gas nebulae surrounding two AGNs, 
whose outflow radii have been constrained by UV absorption line analyses.
In Mrk 509, we find a quasi-spherical outflow with a radius of 
1.2 kpc and a velocity of $\sim290$ km s$^{-1}$, while 
IRAS F04250$-$5718 is driving a biconical outflow extending out to 2.9 kpc, 
with a velocity of $\sim580$ km s$^{-1}$ and an opening angle of $\sim70\degr$.
The derived mass flow rate is $\sim5$ and $>1$ M$_{\odot}$ yr$^{-1}$, respectively, 
and the kinetic luminosity is $\gtrsim1\times10^{41}$ erg s$^{-1}$ for both.
Adopting the outflow radii and geometric parameters measured from IFU, absorption
line analyses would yield mass flow rates and kinetic luminosities in agreement with 
the above results within a factor of $\sim2$.
We conclude that the spatial locations, kinematics and energetics revealed by this IFU 
emission-line study are consistent with pre-existing UV absorption line 
analyses, providing a long-awaited direct confirmation of the latter as an effective 
approach for characterizing outflow properties.

\end{abstract}





\section{INTRODUCTION}
\label{sec:intro}

The active phase of the super-massive black holes residing in galaxy centers has
crucial effects on the global properties of their host galaxies. Thus, feedback
from active galactic nuclei (AGNs) has become an indispensable ingredient of galaxy 
formation models, requiring the black holes to blow large-scale outflowing winds 
which clear out the surrounding gas and prevent future star formation 
\citep[e.g.][]{Tabor93,DiMatteo05,Hopkins10}.
Roughly 20--40\% of quasars show blueshifted broad absorption lines (BAL), 
implying that sub-relativistic quasar outflows are ubiquitous 
\citep{Hewett03,Ganguly08,Knigge08,Dai08}.
Theoretical modeling has shown that massive, wide angle outflows can 
indeed be very efficient feedback agents 
\citep[e.g.][]{Ostriker10,Ciotti10,Soker10,McCarthy10,Hopkins10,FaucherGiguere12,Choi14}. 

Over the last decade, our collaboration and other research groups have been performing 
absorption line analyses on about 20 AGN absorption outflows to measure the distance
from the central source, the mass flow rate and kinetic luminosity 
\citep{Hamann01,deKool01,deKool02a,deKool02b,Gabel05,Moe09,Bautista10,
Dunn10,Aoki11,Arav12,Borguet12a,Borguet12b,Borguet13,Edmonds11,Arav13,Lucy14,Chamberlain15}. 
Our state-of-the-art analysis methods are able to determine the outflow sizes
to better than a factor of two \citep{Bautista10}, frequently finding them located at kpc 
scales (see \citealt{Arav13} for a review), comparable to the spatial 
extent of the host galaxies. The main weakness of these investigations, obtained by 
ultraviolet (UV) spectroscopy of excited transitions coupled with detailed photoionization 
modeling, is the lack of spatial information. The indirectly inferred outflow sizes remain
to be confirmed, not to mention their morphology and spatial structure.

In recent years, integral field unit (IFU) observations have been conducted to map the ionized gas 
surrounding AGNs \citep[e.g.][]{Nesvadba08,Husemann08,Fu09,Lipari09,Alexander10,Humphrey10,Rupke11,
Husemann11,CanoDiaz12,Harrison12,Westmoquette12,Husemann13,Liu13a,Liu13b,Rupke13,Harrison14,Liu14,
Shih14,McElroy15,Davies15}. In particular, this technique has enabled us to detect massive, 
powerful, kpc-scale, AGN-driven outflows both in radio-quiet major mergers \citep{Rupke11,Rupke13}
and in high luminosity radio-quiet type 2 \citep{Liu13a,Liu13b} and type 1 \citep{Liu14} quasars, 
providing evidence for ubiquitous galaxy-wide feedback in luminous quasars.

IFU spectroscopy has the obvious advantage of directly measuring the spatial extent of the 
outflows which can only be derived indirectly from absorption line analyses (not to mention 
the morphology and the spatial distribution of gas kinematics that can only be obtained with IFU),
yet the latter provides more abundant spectroscopic information covering a wide range of ionization 
and density, enabling the probe of the physical conditions in the multi-phase outflowing gas \citep{Arav13}. 
Therefore, IFU mapping of outflowing AGNs will directly test the outflow locations derived 
from UV absorption lines, providing a strong constraint on the physical assumptions and parameters 
of the outflow models.

In this paper, we present Gemini IFU mapping of the ionized gas nebulae around two bright radio-quiet
active galaxies, Mrk 509 and IRAS F04250$-$5718. The locations of their outflows both have been determined or constrained by 
previous UV absorption line analyses using high-quality spectroscopy.

Although spectroscopically classified as Seyfert 1.5 galaxies (Table \ref{tab1}), Mrk 509 and 
IRAS F04250$-$5718 both reside close to the quasar/Seyfert galaxy borderline \citep{Kopylov74,VeronCetty06}, 
and are thus deemed low-luminosity quasars.
As one of the best studied AGNs, Mrk 509 ($z=0.03440$ \citealt{Huchra93}) has been scrutinized
in multiple windows of its electromagnetic spectrum (a thorough description of its observational properties
is available in \citealt{Kaastra11}). Combining HST/COS data from a massive multi-wavelength monitoring 
campaign \citep[see][]{Kaastra11} and archival HST/STIS data, \citet{Arav12} analyzed all the kinematic 
components of its outflow and obtained a lower limit of 100--200 pc for their galactocentric distance.

IRAS F04250$-$5718 ($z=0.104$, \citealt{Thomas98}) was discovered in X-ray as 1H 0419$-$577 \citealt{Wood84},
and is also known as LB 1727 and 1ES 0425$-$573. So far, the X-ray spectral energy distribution has been the 
primary interest of the existing studies on it (the reader is referred to the introduction section of 
\citet{Edmonds11} for a detailed description of its observational history). 
\citet{Edmonds11} identified three kinematic components from its HST/COS UV spectra, and 
determined a conservative lower limit of 3 kpc for the outflow. The basic characteristics of these two
objects and the observing information of our observations are summarized in Table \ref{tab1}.

This paper is organized as follows. In Section \ref{sec:data}, we describe the observations and reduction
of our data. Flux calibration, PSF subtraction and line fitting are presented
in Section \ref{sec:analysis}. Outflow features are identified and characterized in Section 
\ref{sec:outflow}. In Section \ref{sec:compare}, we measure the spatial scales and derive
physical velocities of the outflows, and compare to absorption line analyses. The mass flow rates and kinetic
luminosities are estimated in Section \ref{sec:discussion}, followed by a summary in Section 
\ref{sec:summary}.
We adopt an $h=0.70$, $\Omega_{\rm m}=0.29$, $\Omega_{\Lambda}=0.71$ cosmology throughout this paper.




\begin{center}
\begin{deluxetable*}{l c r c r c c c l}[ht]
\setlength{\tabcolsep}{0.02in} 
\tablecaption{Characteristics of the sample objects and observational parameters. \label{tab1}}
\tablehead{
\colhead{Object} & \colhead{$\alpha,\;\delta$ (J2000.0)} & \colhead{$z$} & \colhead{Type} & \colhead{PA} 
& \colhead{$t_{\rm exp}$} & \colhead{Seeing} & \colhead{$L_{\rm [O\;{\scriptscriptstyle III}]}$} & \colhead{$\Sigma_{\rm [O\;{\scriptscriptstyle III}],peak}$}\\ 
\colhead{(1)} & \colhead{(2)} & \colhead{(3)} & \colhead{(4)} & \colhead{(5)} 
& \colhead{(6)} & \colhead{(7)} & \colhead{(8)} & \colhead{(9)} 
}

\startdata

Mrk 509             &  $20^{\rm h}44^{\rm m}09\fs7\,-\!10\degr43\arcmin24\farcs5$ &  0.0344   &  Seyfert 1.5 &  0\degr & 1200$\times$2 & 0\farcs70 & $1.9\!\times\!10^{42}$  & $4.9\!\times\!10^{-13}$   \\  
IRAS F04250$-$5718  &  $04^{\rm h}26^{\rm m}00\fs7\,-\!57\degr12\arcmin01\farcs0$ &  0.104    &  Seyfert 1.5 & 90\degr & 1600$\times$2 & 0\farcs62 & $1.4\!\times\!10^{43}$  & $2.5\!\times\!10^{-13}$


\enddata

\tablecomments{
(1) Object name. 
(2) Right ascension and declination.
(3) Redshifts are taken from \citet[][for IRAS F04250$-$5718]{Thomas98} and \citet[][for Mrk 509]{Huchra93}.
(4) AGN type, from the NASA/IPAC Extragalactic Database (NED, http://ned.ipac.caltech.edu).
(5) Position angle of the field of view, measured by the longer (5\arcsec) axis of the FoV (1-slit IFU) from north to east.
(6) Exposure time (in seconds) and number of exposures.
(7) Seeing at the observing site (FWHM).
(8) Total luminosity of the \oiii$\lambda$5007 \AA\ line (erg s$^{-1}$), from observations by
\citet[][for IRAS F04250$-$5718]{Guainazzi98} and \citet[][for Mrk 509]{Carone96}.
(9) Peak \oiii$\lambda$5007 \AA\ surface brightness before PSF subtraction (erg s$^{-1}$ cm$^{-2}$ arcsec$^{-2}$).
}
\end{deluxetable*}
\end{center}


\section{Observations and data reduction}
\label{sec:data}

We observed the program targets with the Gemini-South telescope between 
2013 July and 2014 January (program ID: GS-2013B-Q-84, PI: D. Rupke;
we also observed IRAS F22456$-$5125 in our campaign, but the data quality 
is insufficient for deriving useful results). We used the 1-slit 
Integral Field Unit (IFU) mode on GMOS-S, so that the wavelength coverage is optimized. 
We performed two exposures, separated by a spatial dithering of 1.5\arcsec\ along the shorter
axis of the field of view (FoV). Combining the dithered exposures, we cover a final FoV
of 5\arcsec$\times$5\arcsec, translating to a linear scale of 3.5$\times$3.5 kpc$^2$ for 
Mrk 509 and 9.5$\times$9.5 kpc$^2$ for IRAS F04250$-$5718. The science FoV, sampled 
by 1000 contiguous 0.2\arcsec-diameter hexagonal lenslets, is interpolated to a finer
grid (0.1\arcsec) in the final data products.

The seeing at the observing site was 0.6\arcsec--0.7\arcsec, measured from the point 
spread function (PSF) model we construct from the continuum emission (see Section \ref{sec:psf}).
We used the spectroscopy blocking filter GG455-G0329 in our program. 
The central wavelength of the employed grating B600-G5323 was
set to 625 nm for IRAS F04250$-$5718 and 615 nm for Mrk 509, respectively.
This grating has a spectral resolution of $R\sim2300$ at these wavelengths.
Both objects were observed between 4700\AA\ and 7600\AA\ so as to cover the \oiii-H$\beta$ region. To 
accurately determine the spectral resolution, we perform Gaussian fits on unresolved sky 
lines in the \oiii-H$\beta$ neighborhood, finding their full width at half maximum 
(FWHM) to be $\sim120$ km s$^{-1}$. The
emission lines of interest (\oiii$\lambda\lambda$5007, 4959\AA, H$\beta$, 
H$\alpha$, [N {\sc ii}]$\lambda\lambda$6548, 6583\AA) are all well 
resolved. Table \ref{tab1} summarizes our Gemini observations.

We use the Gemini package for IRAF\footnote{The Image Reduction and Analysis Facility (IRAF) is 
distributed by the National Optical Astronomy Observatories which is operated 
by the Association of Universities for Research in Astronomy, Inc. under 
cooperative agreement with the National Science Foundation.} to reduce the data, 
following the standard procedure for GMOS IFU described in the tasks {\sl gmosinfoifu} 
and {\sl gmosexamples} 
with slight modification (detailed in \citealt{Liu13a}).
The datacube product reduced from each exposure has 0.1\arcsec\ spatial pixels (``spaxels''). 
We finally combine the two frames using the task {\sl imcombine} by taking the mean spectra 
in each spaxel. The \oiii5007\AA\ line is well detected in every spatial position, with
signal-to-noise ratios at its peak of 10--20 at the FoV edges, and 2500--2800 in the FoV 
centers.

\section{IFU data analysis}
\label{sec:analysis}

\subsection{Flux calibration and continuum / \feii\ subtraction} 
\label{sec:calib}

We flux-calibrate our data against previous spectroscopic measurements. 
For IRAS F04250$-$5718, \citet{Guainazzi98} find an integrated \oiii\ flux of 
$f_{\rm [O \scriptscriptstyle{III}]}=5.0\times10^{-13}$ erg s$^{-1}$ cm$^{-2}$ 
using the ESO 1.52m telescope equipped with an 8\arcsec-wide long slit, while a 
5-year coordinated program of spectroscopic monitoring of Mrk 509 reports  
$f_{\rm [O \scriptscriptstyle{III}]}=6.79\times10^{-13}$ erg s$^{-1}$ cm$^{-2}$ 
using various apertures sizes, of which the smallest is $5\arcsec\times8\arcsec$
\citep{Carone96}. Flux loss from these measurements are deemed negligible 
because the employed apertures are larger than our field of view. 

In order to remove the contamination from \feii\ emission, we perform a
procedure identical to \citet{Liu14}. We fit the continua in the 
\oiii-H$\beta$ neighborhood with the sum of two components:
a quadratic polynomial, and the \feii\ template from \citet{Boroson92}
smoothed using a Gaussian kernel whose width is one of the free fitting 
parameters. 
However, we note that \feii\ contamination to our \oiii\ analysis is insignificant 
in both targets. 

\subsection{PSF subtraction}
\label{sec:psf}

Our program was undertaken when seeing at the observing site was  
0.6\arcsec--0.7\arcsec, requiring careful analysis of our datasets to minimize the
effects of the PSF. The observations become significantly easier to interpret
if the PSF is carefully modeled and subtracted from the IFU datacube. The 
most reliable approach is constructing PSF models from the real
data themselves. 

Quasar light scattered by the interstellar matter \citep{Borguet08} or star 
formation in the quasar host \citep{Letawe07,Silverman09} might both contribute
to the continuum emission. However, we assume that the continuum emission 
is dominated by a central point source in this work. This is a reasonable 
assumption for both of our targets that are classified as type 1 AGNs, and 
is further validated by the fact that the FoV is only larger than the FWHM 
of the PSF by a factor of 6, and thus the scattered light from the host 
galaxy, if detectable at all in the outer regions, is overwhelmed by the 
broad wings of the PSF. 

To construct a PSF model in an optimized way, we first select three rest-frame 
wavelength intervals free of \oiii\ and H$\beta$ line emission, 4925--4945\AA,
4970--4980\AA\ and 5020--5050\AA, normalize the central brightest pixel
to unity, then create a median image for each interval. For where \oiii\ and 
H$\beta$ emission exist, we linearly interpolate between each pair of
adjacent median images to construct the PSF model at each wavelength, so that 
a PSF model datacube is created. These constructed PSFs have an FWHM of 
0.62\arcsec\ for IRAS F04250$-$5718 and 0.70\arcsec\ for Mrk 509.

For the \oiii--H$\beta$ region, we perform Gaussian fits to both the PSF 
model at each wavelength and to the corresponding slice image to compute 
their centroids using the {\sc gcntrd} routine in the IDL Astronomy User's 
Library\footnote{http://idlastro.gsfc.nasa.gov/} (Gaussian fits are 
sufficient for our purpose of centroiding only). We then align the PSF
to the slice image using the obtained centroids (although the position of 
the AGN is fixed in the sky, we allow the peak of the PSF to move minimally 
in the plane of the sky (within one spatial pixel, 0.1\arcsec) for optimized 
results). After that, the PSF is scaled to the central peak brightness of 
the slice image, and subtracted from the latter. As the last step, guided by 
these automated algorithms, we manually adjust the brightness of the 
PSF to further correct imperfect subtractions.

\subsection{Spectral line fits}

After removing Fe {\sc ii} emission and subtracting the PSF from the IFU
datacubes, we perform multi-Gaussian fits to the [O {\sc iii}]5007,4959
doublet, so that a noiseless model of the line in every spatial pixel is
attained, a procedure identical to \citet{Liu13b}. As described in
that paper, up to 3 Gaussian components are needed for good fits so that 
the reduced $\chi^2\lesssim 2$ in every spatial pixel, and the
actual number of employed Gaussians is determined by comparing their 
respective reduced $\chi^2$ values. We then compute the \oiii\ line 
intensity in every spatial position from the multi-Gaussian fit (instead
of the observed profile). The surface brightness sensitivity 
of our \oiii\ surface brightness maps is measured to be 
$\sigma_{\rm [O\;{\scriptscriptstyle III}]}=(3$--$6)\times 10^{-17}$ 
erg s$^{-1}$ cm$^{-2}$ arcsec$^{-2}$.

As in \citet{Liu13b}, we do not assign 
physical meanings to the individual Gaussians (with the sole exception
of decomposing \oiii\ to remove the tidal tail emission in Mrk 509,
see Section \ref{sec:M509outflow}). The fits are merely used to produce
noiseless models of the \oiii\ line to facilitates non-parametric 
measurements. 

Following \citet{Whittle85} and \citet{Liu13b}, we measure two quantities
that characterize the line-of-sight velocity and velocity dispersion of the
ionized gas, the median velocity ($v_{\rm med}$) that bisects the total
area underneath the \oiii\ emission line profile, and the velocity interval 
that encloses 80\% of the total \oiii\ emission centered at the median 
velocity ($W_{80}$). As noted in those papers, $W_{80}$ is more sensitive
to the weak broad wings of a non-Gaussian profile, but is similar to FWHM
for a Gaussian velocity profile ($W_{80}=1.088\times {\rm FWHM}$).

\section{Outflow signatures}
\label{sec:outflow}

\subsection{IRAS F04250--5718}

\begin{figure*}
\includegraphics[scale=0.33,clip=true,trim=0cm 0mm 10mm 0mm]{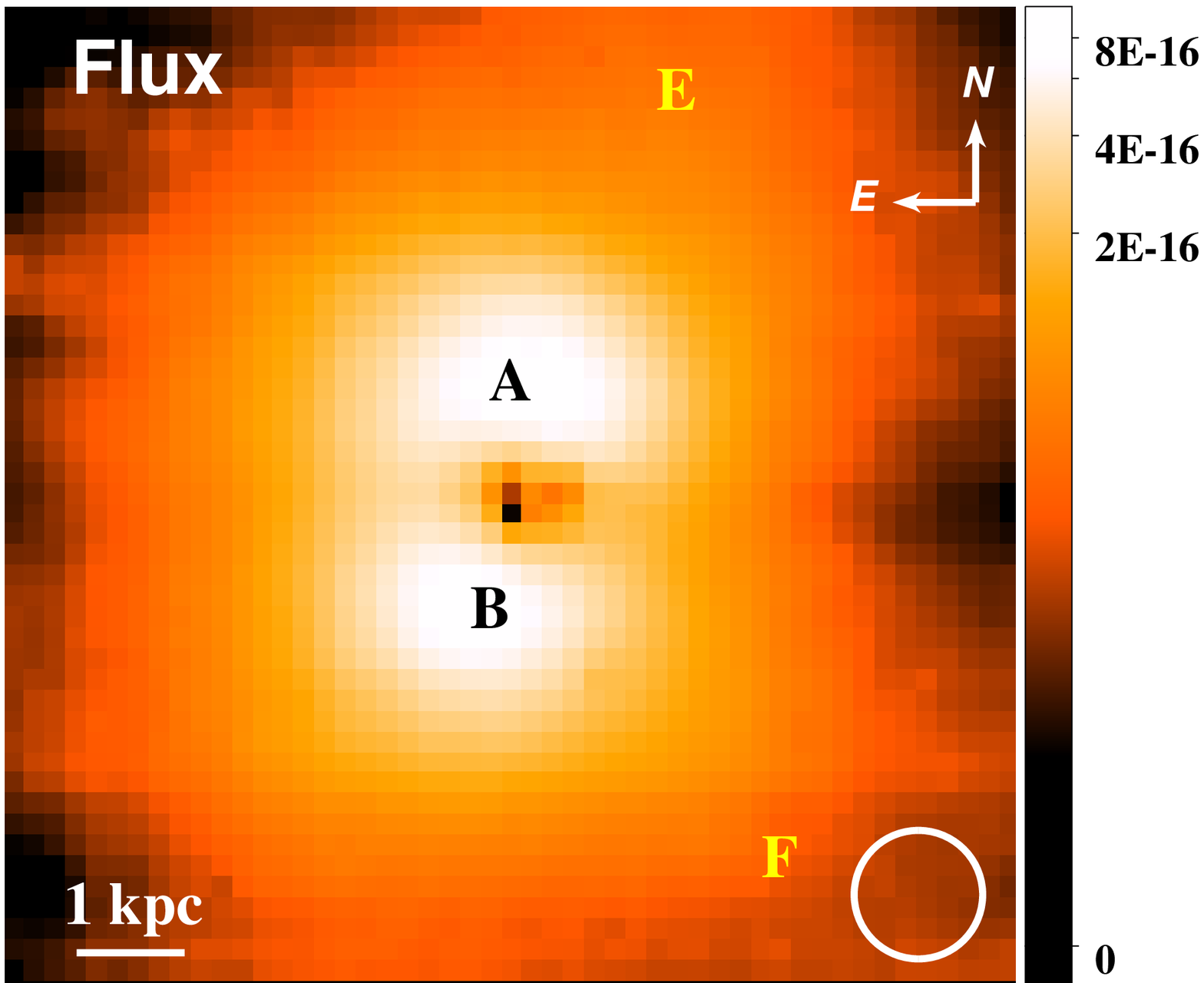}%
\includegraphics[scale=0.33,clip=true,trim=0cm 0mm 17mm 0mm]{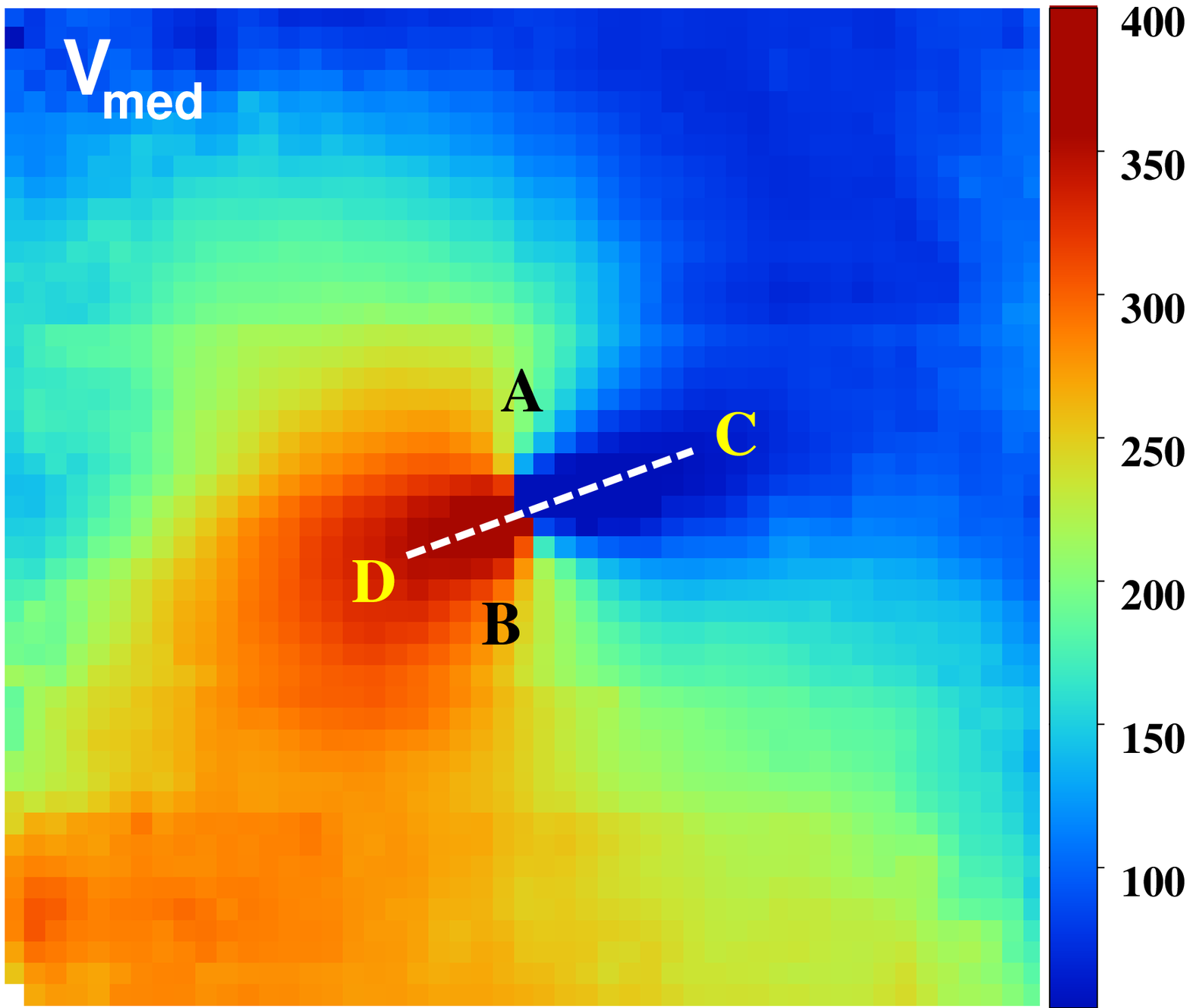}%
\includegraphics[scale=0.33,trim=0cm 0mm 17mm 0mm,clip=true]{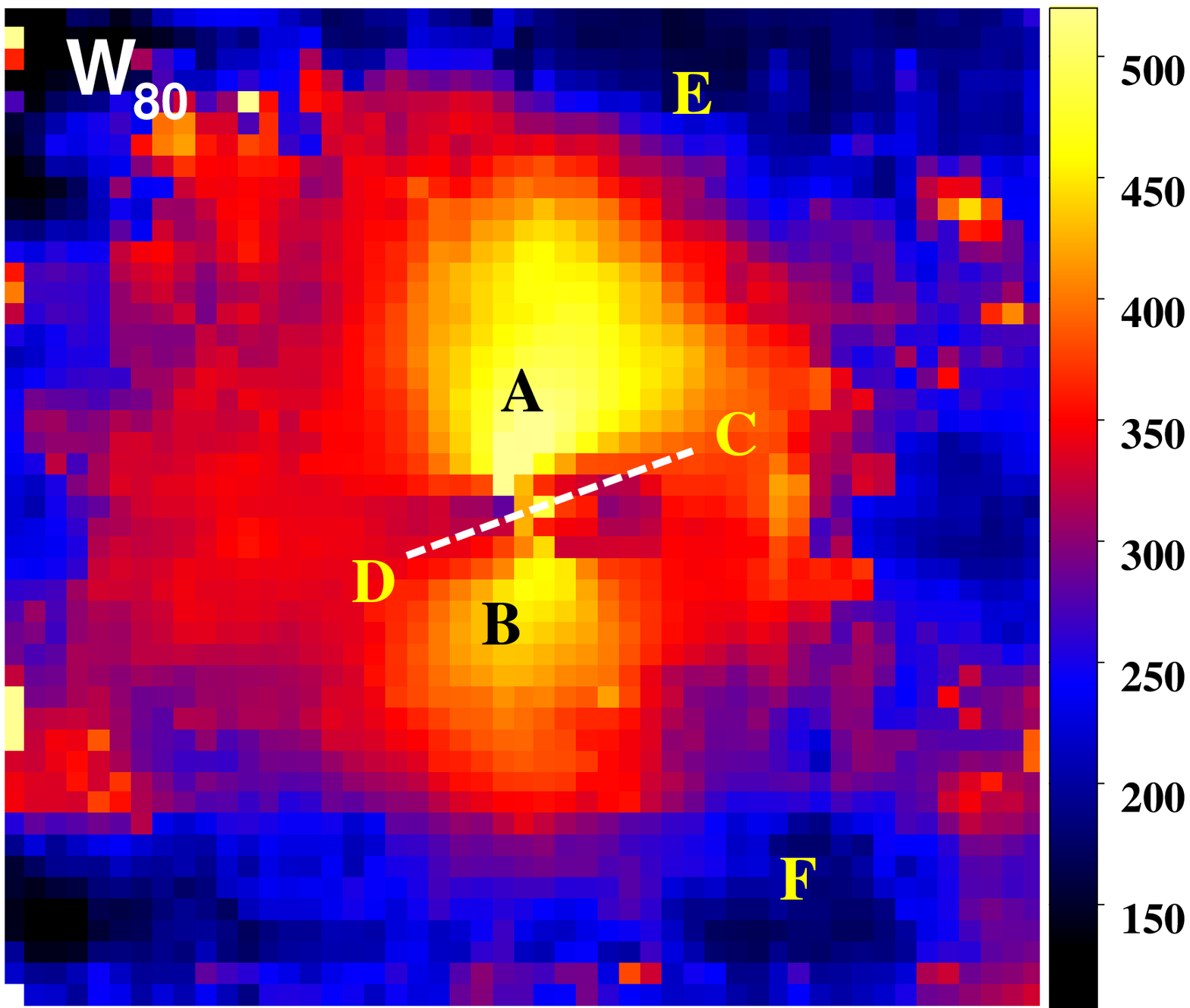}\\
\includegraphics[scale=0.37,clip=true,trim=0cm 20mm 0mm 0mm]{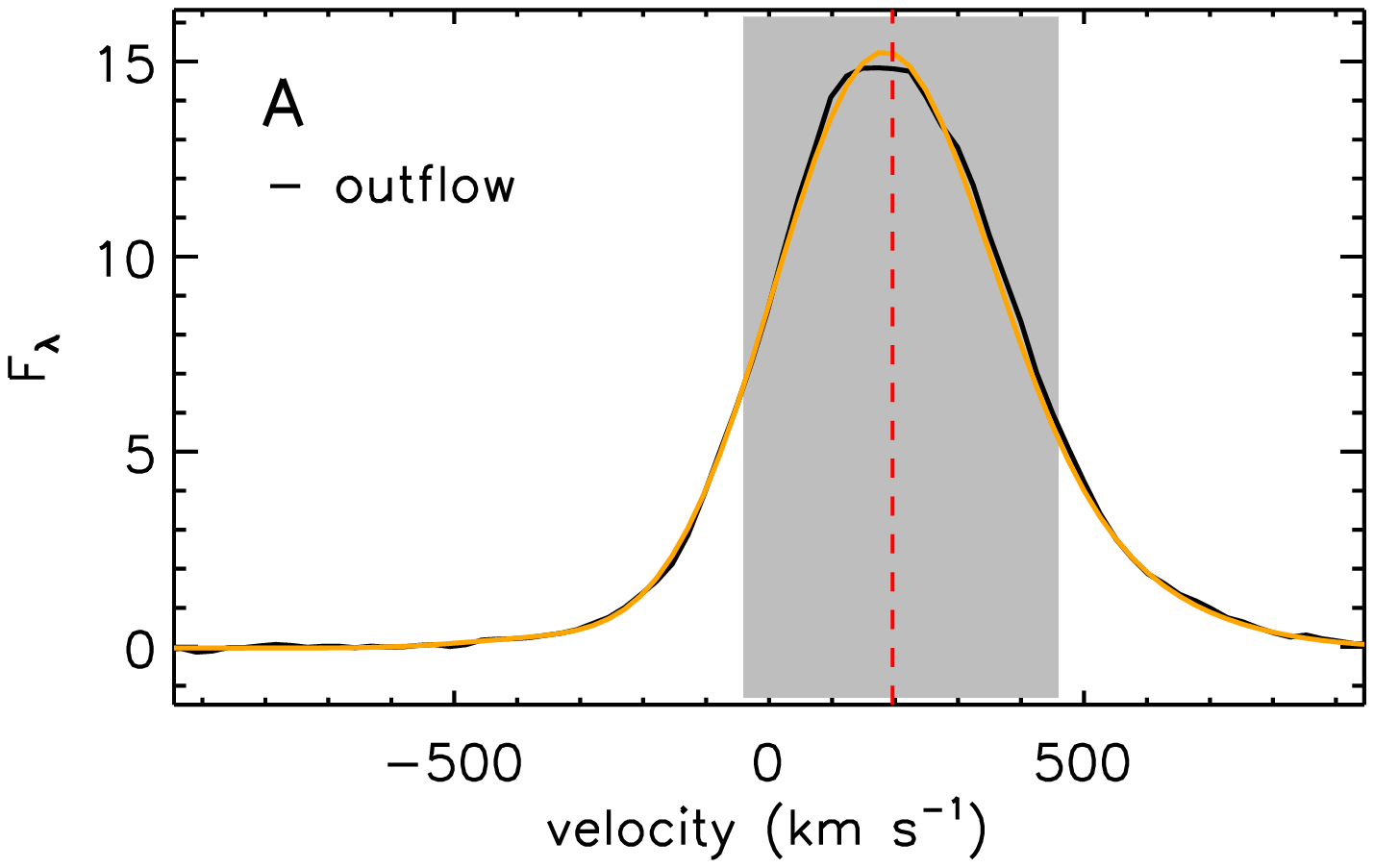}
\includegraphics[scale=0.37,clip=true,trim=0cm 20mm 0mm 0mm]{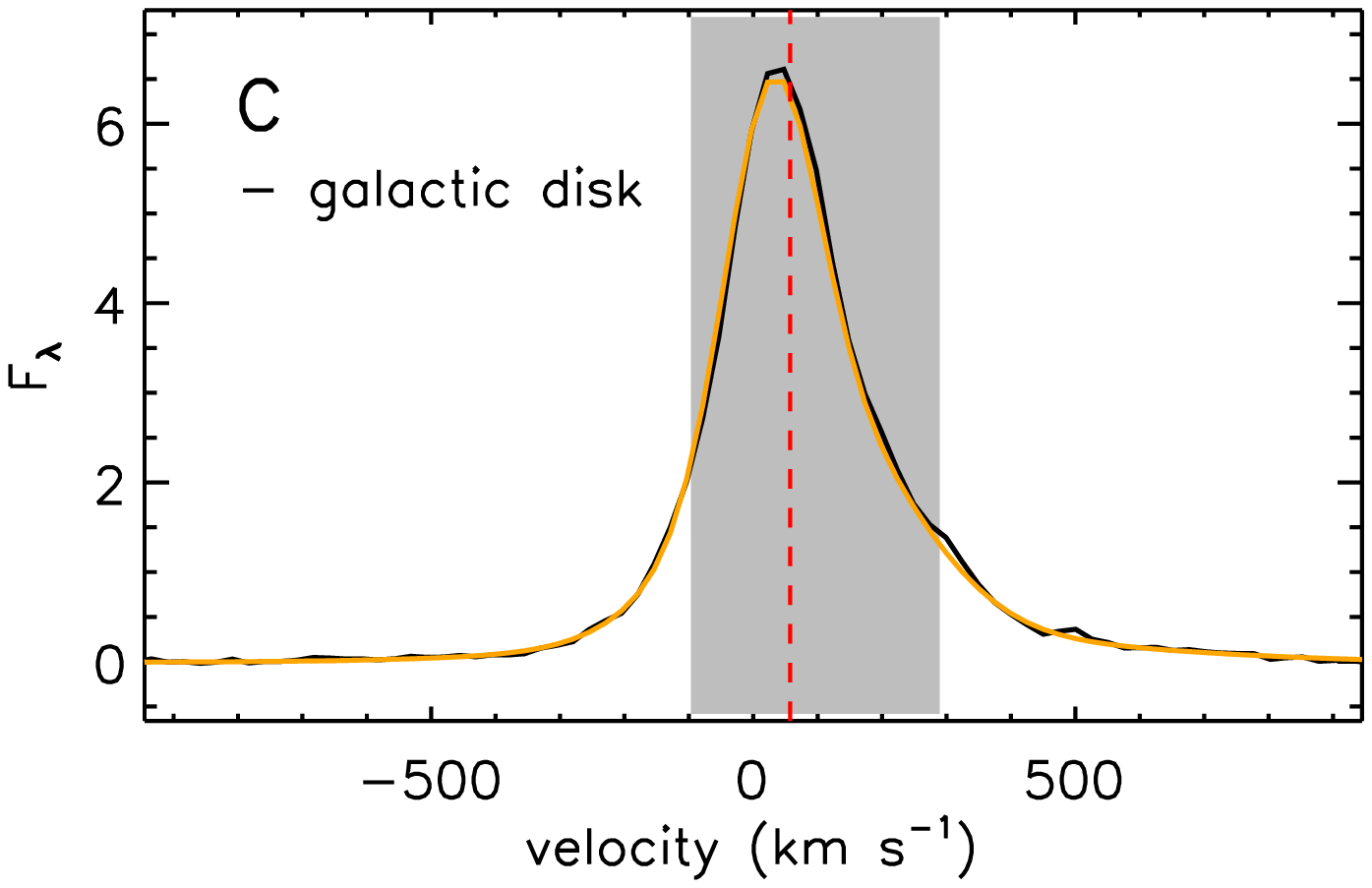}
\includegraphics[scale=0.37,clip=true,trim=0cm 20mm 0mm 0mm]{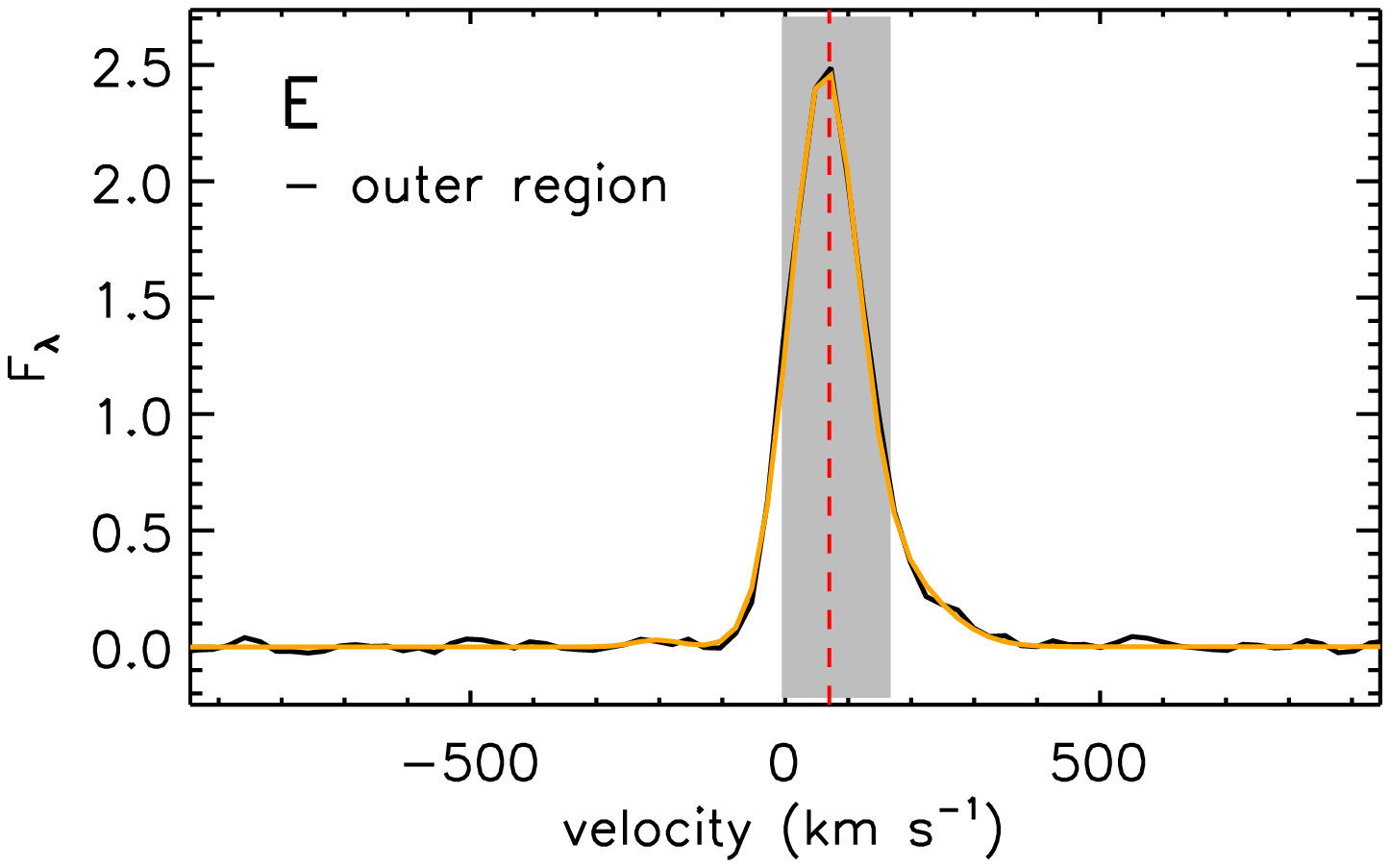}\\
\includegraphics[scale=0.37,clip=true,trim=0cm 0mm 0mm 5mm]{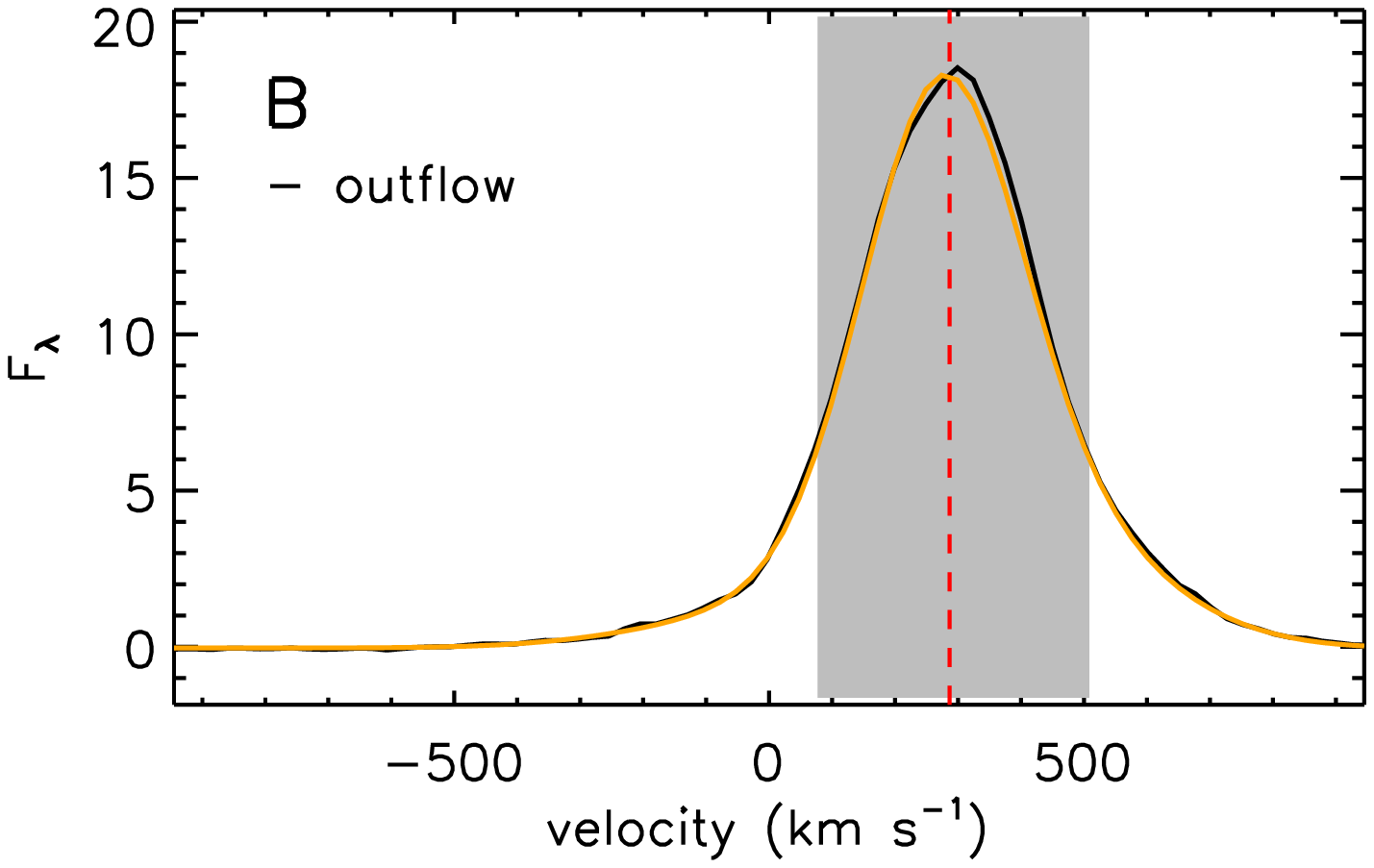}
\includegraphics[scale=0.37,clip=true,trim=0cm 0mm 0mm 5mm]{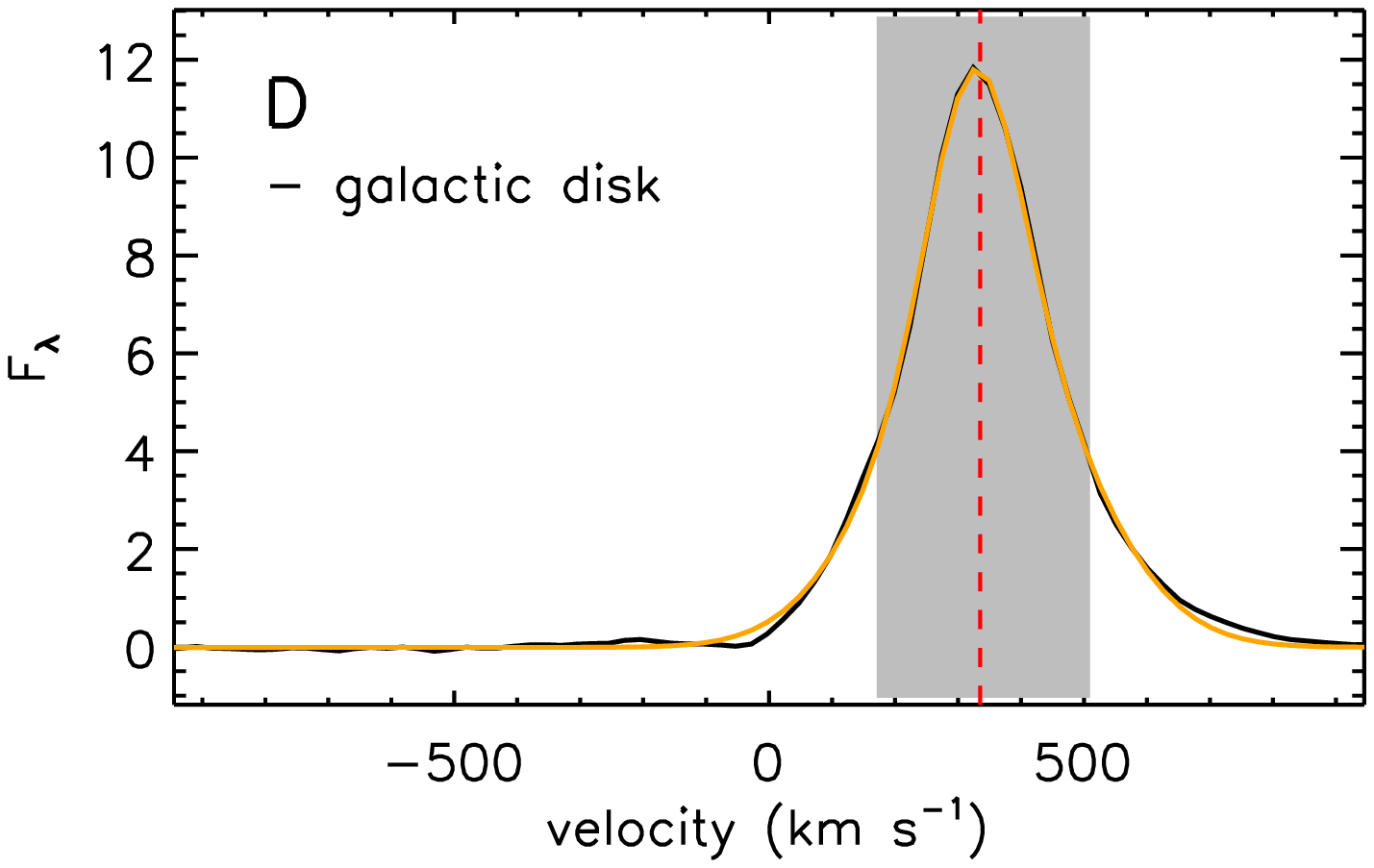}
\includegraphics[scale=0.37,clip=true,trim=0cm 0mm 0mm 5mm]{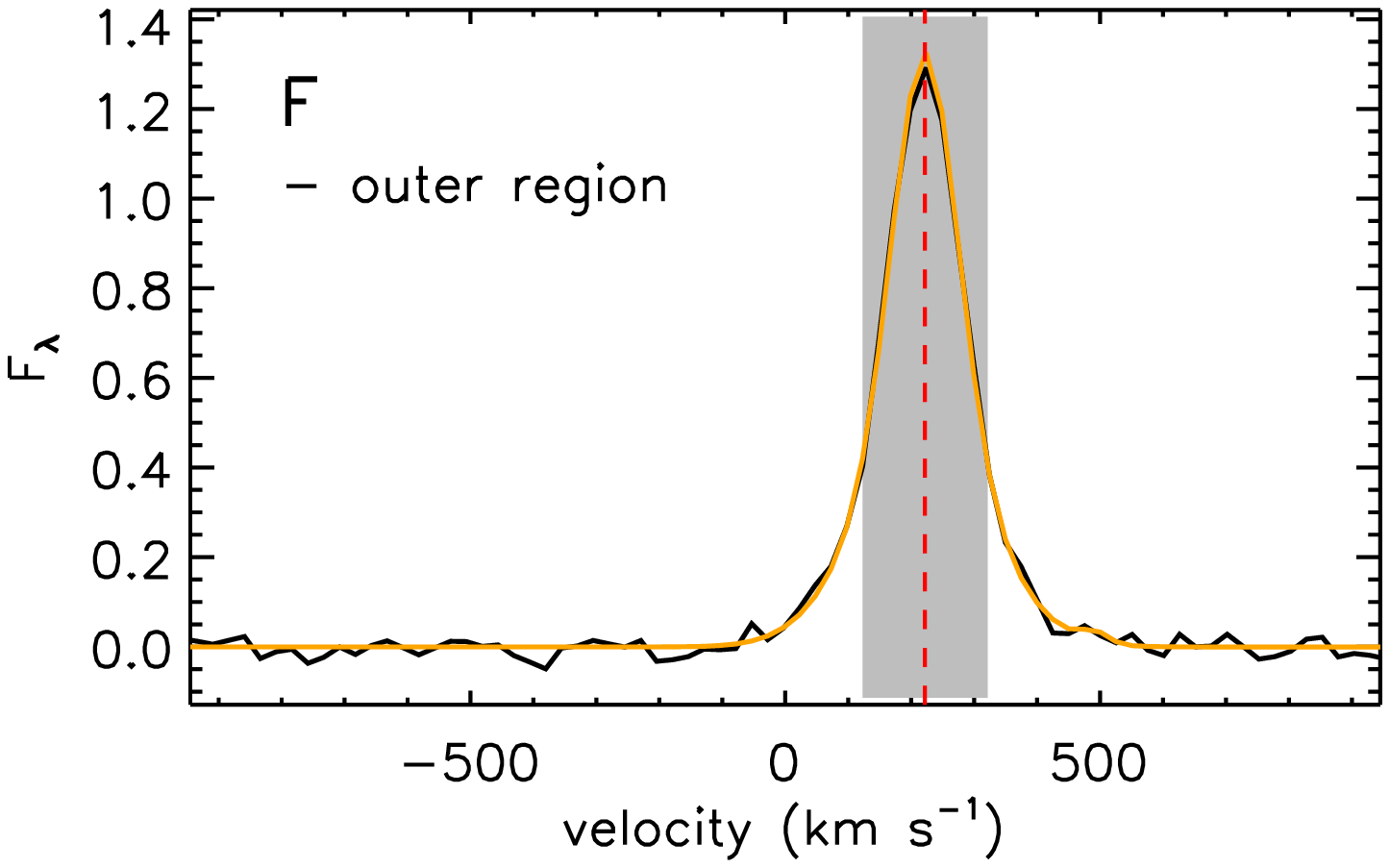}
\caption{IRAS F04250$-$5718. 
{\bf Top color maps.} Maps of the \oiii$\lambda$5007\AA\ surface brightness (left,
in units of erg s$^{-1}$ cm$^{-2}$ arcsec$^{-2}$, logarithmic scale), 
median velocity (middle, in km s$^{-1}$) and line width ($W_{80}$, right,
in km s$^{-1}$). The seeing at the observing site is depicted by the open circle on the surface 
brightness map. The white dashed line on the $V_{\rm med}$ and $W_{80}$ maps
marks the orientation of the nearly edge-on galactic disk rotating at $\sim170$ km s$^{-1}$.
{\bf Bottom spectra.} We select 6 spatial positions to present the \oiii\ velocity 
profile therein ($F_{\lambda}$ is in units of $10^{-17}$ erg s$^{-1}$ cm$^{-2}$ \AA$^{-1}$,
within a 0.1\arcsec\ spatial pixel), 
including 2 outflow regions (``A'', ``B''), 2 galactic
disk regions (``C'', ``D''), and two outer regions (``E'', ``F''). The fitted
line is in orange, the median velocity is marked by red dashed lines, and
the velocity range used for calculating $W_{80}$ is denoted by grey boxes.} 
\label{fig:F04}
\end{figure*}

The \oiii\ surface brightness, median velocity and $W_{80}$ maps are shown
in Figure~\ref{fig:F04}. A pair of strong [O {\sc iii}]-emitting blobs reside
in both sides of the central AGN (the central black pixels are due to slight
over-subtraction), featuring the prevalent bi-conical structure of the AGN
ionization in low-redshift active galaxies (e.g. \citealt{Fischer13}). 
The roughly vertical gap between the two blobs therefore marks the distribution 
of obscuring material with opacity higher than inside the \oiii\ blobs. 
This picture is well consistent 
with the median velocity map on which the distribution of its highest and 
lowest velocities clearly depict a rotation pattern remarkably similar to
a highly inclined ($i\gtrsim70\degr$) disk galaxy (to confirm this 
judgement, we use the publicly available software {\sc DiskFit} 
\citep{Spekkens07,Sellwood10,KuziodeNaray12}
to fit the velocity field of the central region to 
a rotating disk model, finding an inclination angle of $i=70\degr\pm10\degr$). 
Half of the maximum velocity difference on the $v_{\rm med}$ map is 170 km s$^{-1}$, 
approximating the line-of-sight rotation speed of a typical edge-on disk 
galaxy, which extends along the direction $\sim20\degr$ from top to the
right.

The $W_{80}$ maps reveal a pair of prominent high-linewidth lobes. This 
structure spatially coincides 
with the off-center brightness peaks seen in the \oiii\ intensity map, and
extends in a direction (nearly horizontal in the figure) $\sim60\degr$ 
from the direction of maximum $W_{\rm med}$ difference which appears to fall
in a low line-width valley (the contours marking approximately the maximum 
and minimum $v_{\rm med}$ values are overlaid on the $W_{80}$ map to guide 
the eyes). The line width of the pair of lobes reach as high as 
$W_{80}\sim$400--500 km s$^{-1}$, strongly pointing to being highly
turbulent outflowing gas (cf. discussion in \citealt{Liu13b}). 

Therefore, the three maps in Figure \ref{fig:F04} are in consistency with 
a simple and familiar physical picture: the AGN resides in the center of a 
nearly edge-on disk galaxy with a rotation speed of $\sim200$ km s$^{-1}$, 
which both obscures its photoionization along the direction of the disk so that 
the surrounding gas forms a wide ($\gtrsim120\degr$) photoionization
bi-cone, and forces the galactic wind to propagate into other directions with
less obstruction. It is unclear why the outflow is tilted
$\sim20\degr$ off the axis of the rotating disk. Possible
explanations include a warped galaxy disk, or a detailed matter distribution 
with further complication that is not revealed by our observations.

High \oiii\ linewidth implies outflowing ionized gas, both because of the
bulk motion of the gas, and more plausibly, because of the broad distribution
of the velocity of the narrow-line clouds embedded in the winds \citep[see
discussion in][]{Liu13b}. In view of the uniform $W_{80}$ value inside the 
galactic disk (spanning a narrow range of 280--320 km s$^{-1}$ over the narrow
region where $V_{\rm med}$ reaches its minimum and maximum, Fig. \ref{fig:F04}), 
we conservatively choose the $W_{80}$=320 km s$^{-1}$ contour as representing 
the boundary of the galactic outflow. The furthest galactocentric distance 
that the outflow reaches is then measured to be 1.5\arcsec, translating to 
2.9 kpc, on the more extended side, and 1.1\arcsec\ or 2.1 kpc on the more
compact side. 

These distances should be deemed as lower limits. The \oiii\ line emission 
is strongly detected in the entire $10\times10$ kpc$^2$ filed of view, thus
the possibility cannot be ruled out that $W_{80}$ falls below 320 km s$^{-1}$ 
when the wind propagates to sufficiently large distances and even become
undetectable. Further more, galactic winds are multi-phase and our observation
only traces the warm ionized gas. A volume-filling, high-temperature, low 
density ionized gas phase has long been proposed \citep{Heckman90}, which
may be physically even more extended and cannot be traced by \oiii\ emission.
Projection effects also lead to underestimating the physical scale if the
outflow does not perfectly lie in the plane of the sky. In spite of these 
uncertainties, the basic conclusion is validated that an outflow with a total 
physical extent of at least 5 kpc is exerting feedback effects on galaxy-wide 
scales.

\subsection{Mrk 509}
\label{sec:M509outflow}

\begin{figure*}
\centering
\includegraphics[scale=0.4,clip=true,trim=0cm 0mm 0mm 0mm]{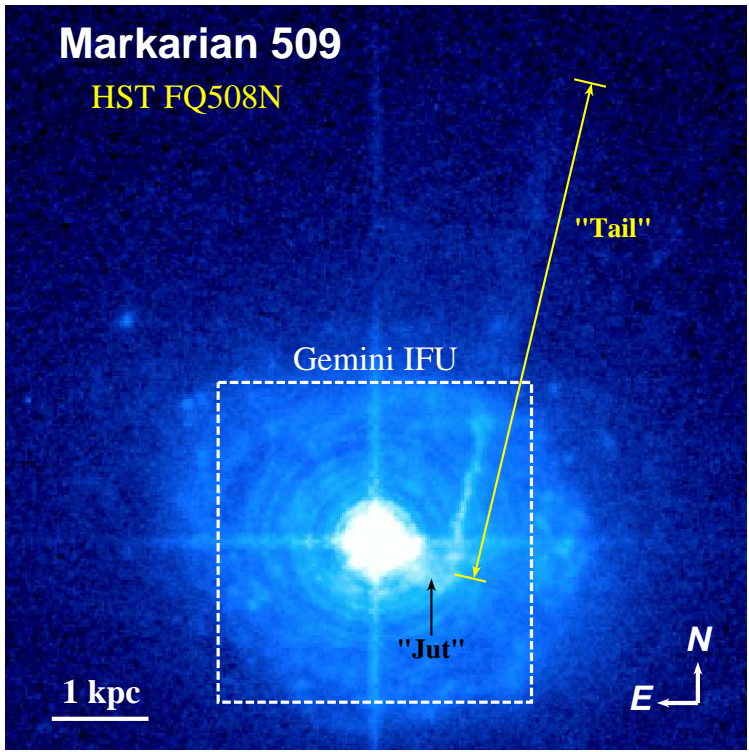}
\vspace*{0.02in}
\caption{HST archival \oiii\ narrow-band image of Mrk 509 (GO--12212; 
PI: D. Crenshaw). The yellow box depicts the 5\arcsec$\times$5\arcsec\ field 
of view of our Gemini IFU observation. The spectacular 5.4-kpc-long linear
``tail'' structure and the ``southwestern jut'' that appears to connect it to 
the center are also marked. See Section \ref{sec:M509outflow} for discussion
of these structures.} 
\label{fig:HST}
\end{figure*}

\begin{figure*}
\centering
\includegraphics[scale=0.28,clip=true,trim=0cm 0mm 10mm 0mm]{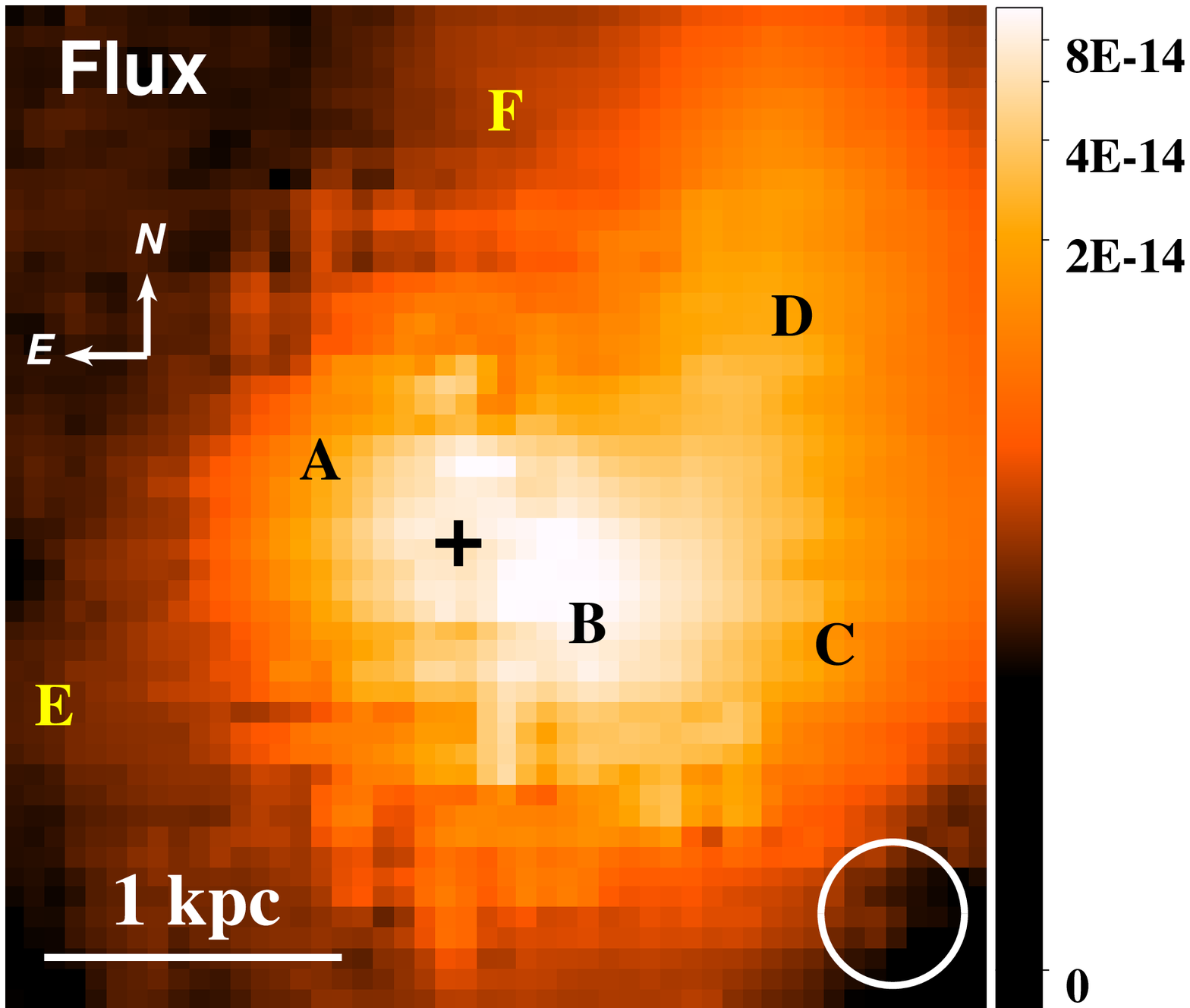}%
\includegraphics[scale=0.28,clip=true,trim=0cm 0mm 18mm 0mm]{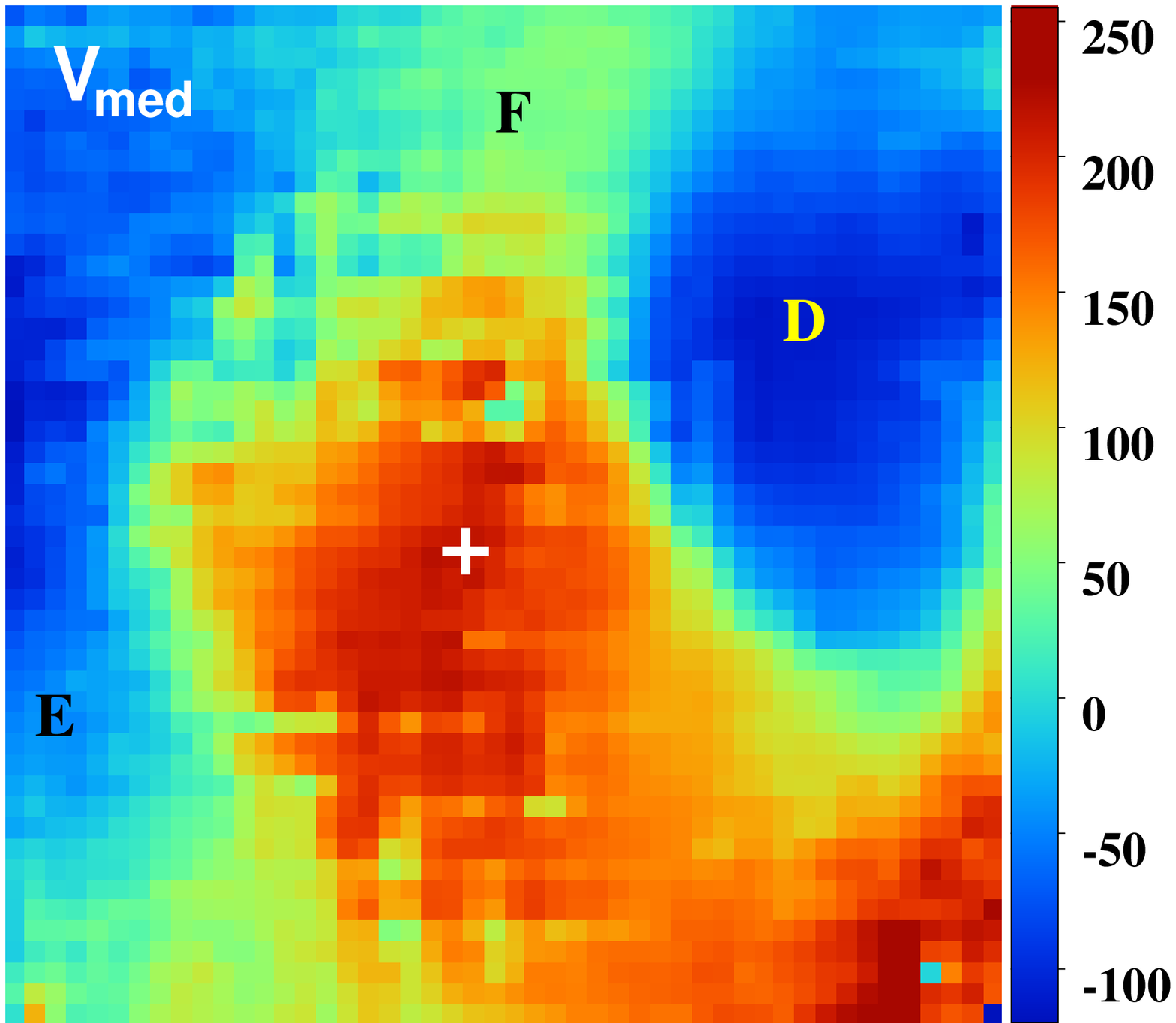}%
\includegraphics[scale=0.28,clip=true,trim=0cm 0mm 25mm 0mm]{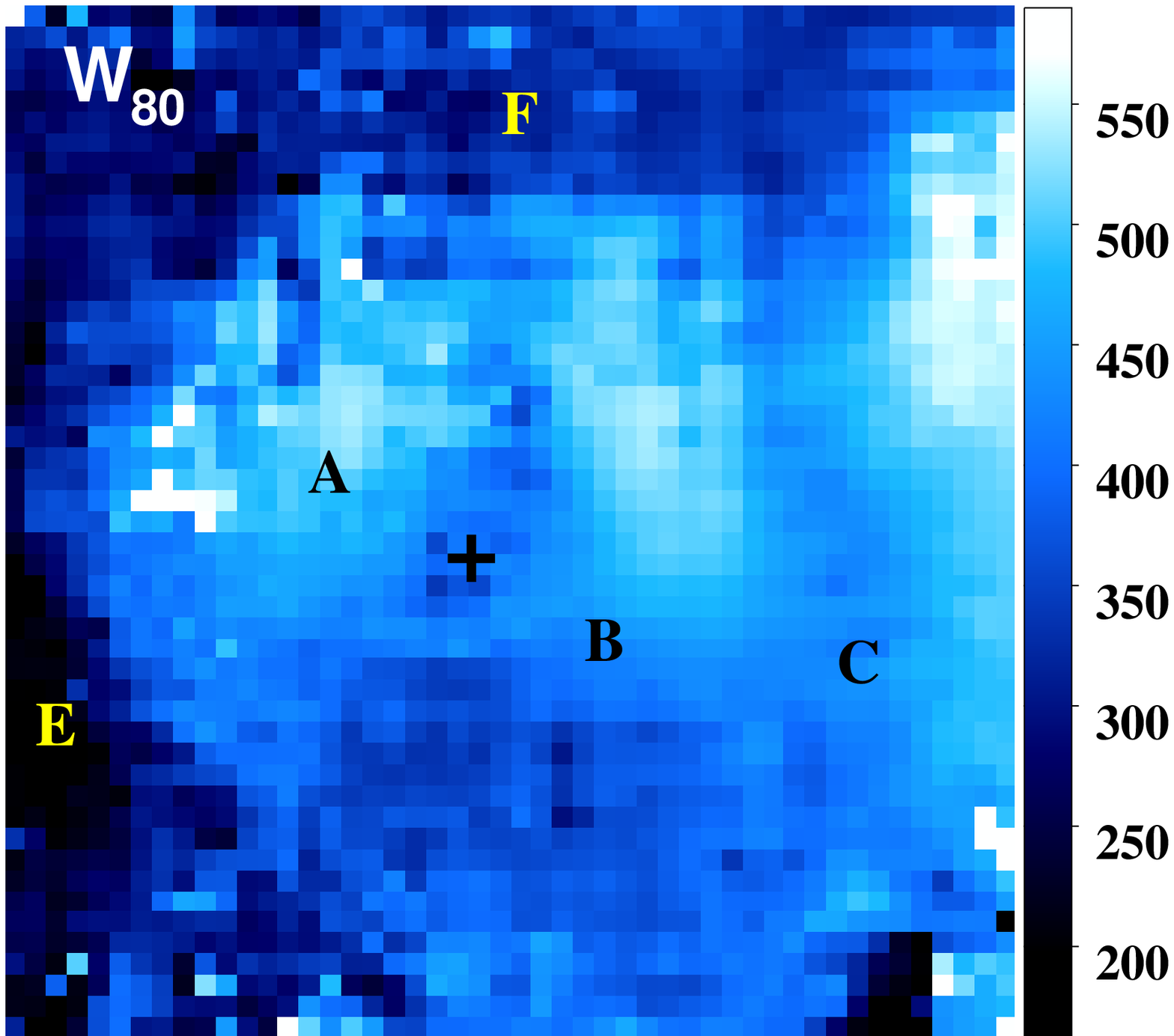}
\includegraphics[scale=0.28,clip=true,trim=0cm 0mm 10mm 0mm]{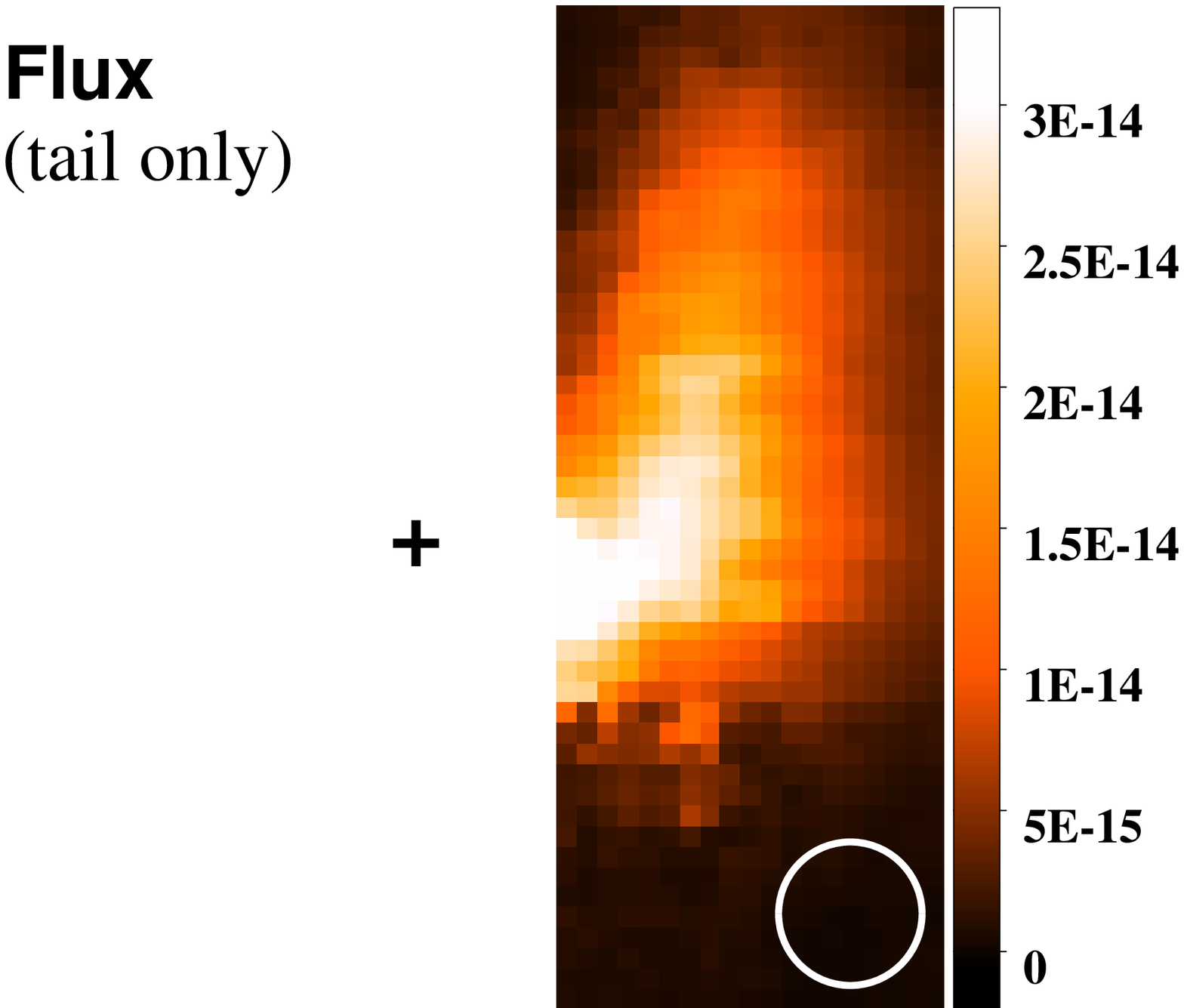}%
\includegraphics[scale=0.28,clip=true,trim=0cm 0mm 18mm 0mm]{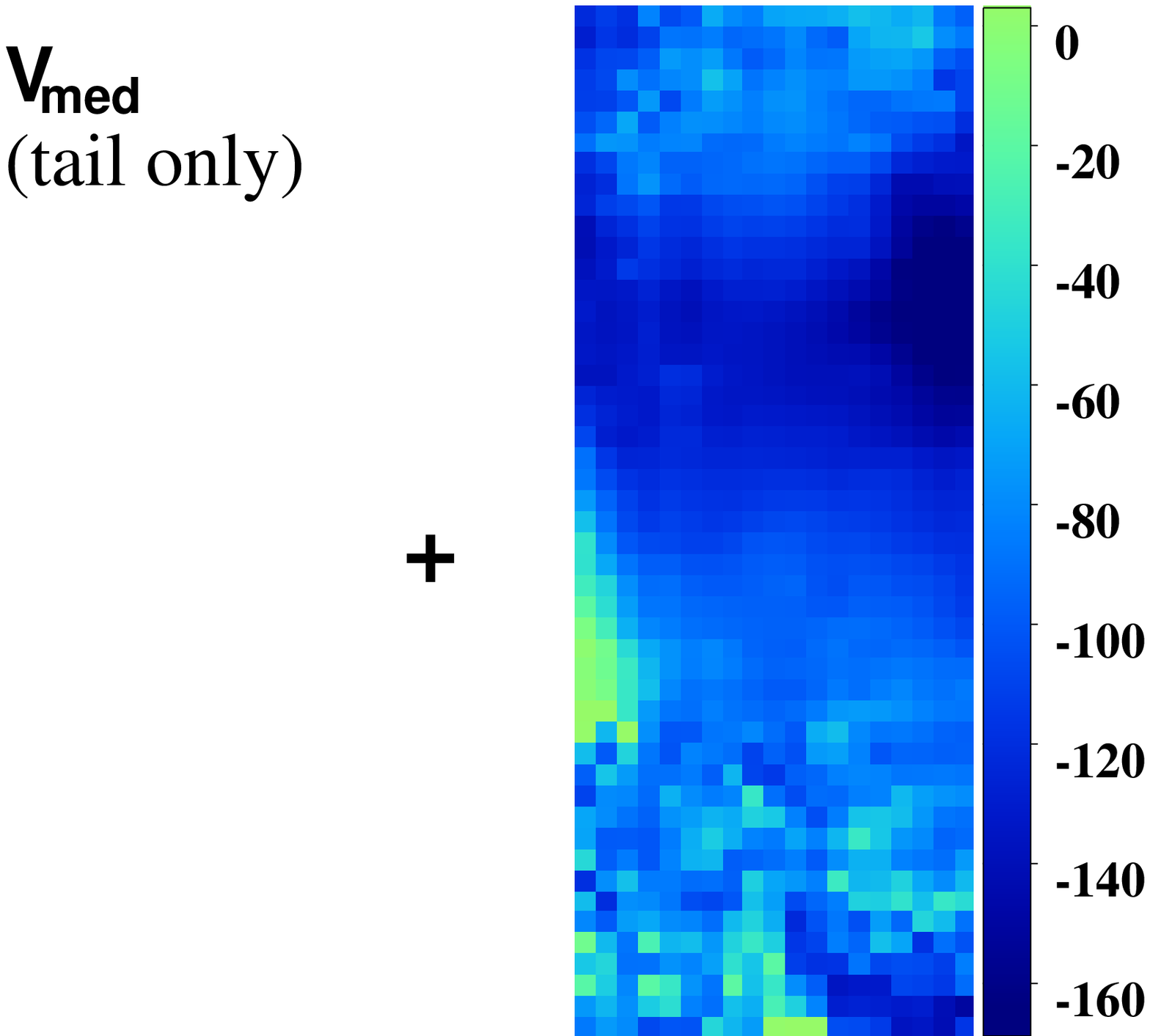}%
\includegraphics[scale=0.28,clip=true,trim=0cm 0mm 25mm 0mm]{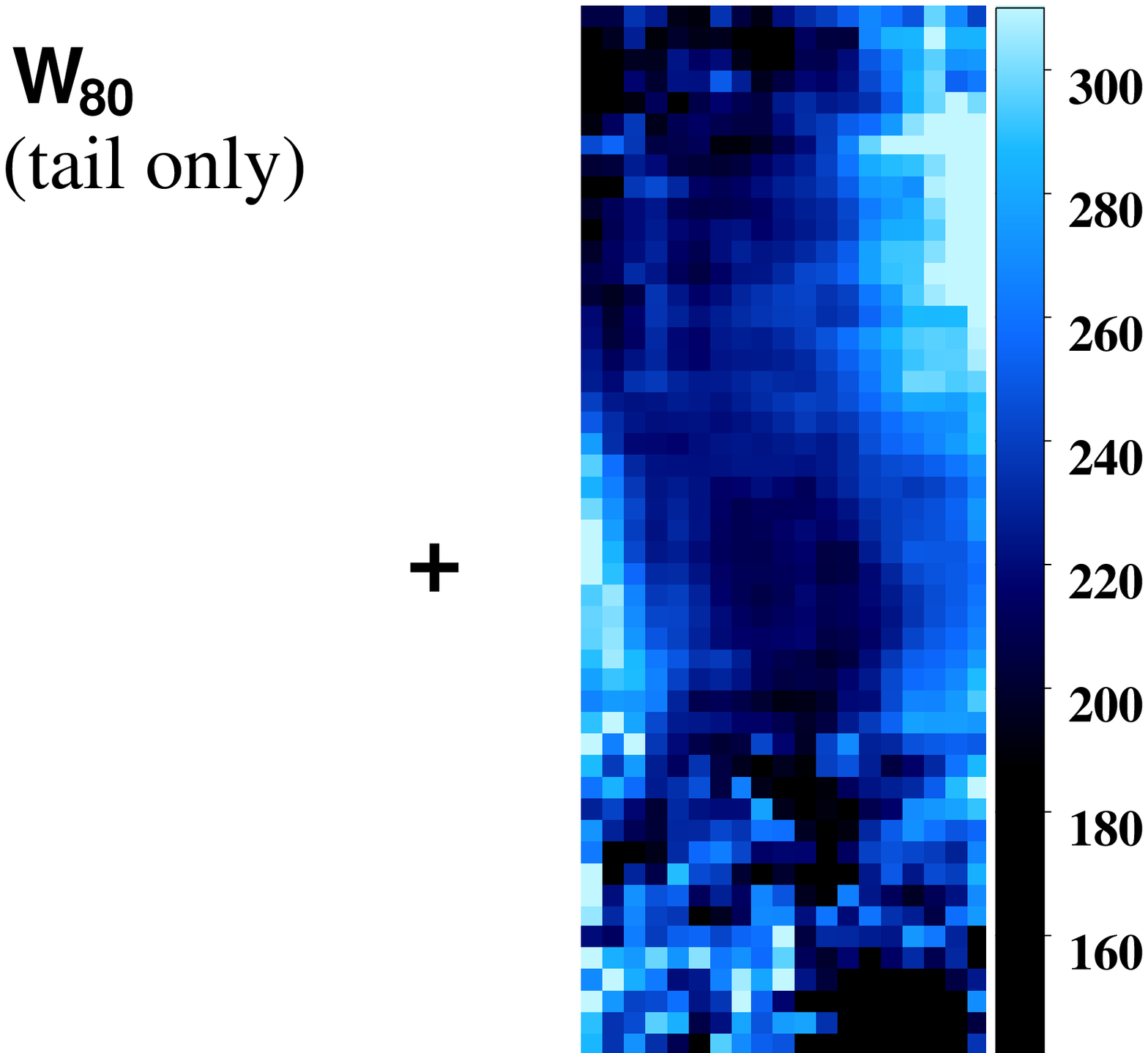}
\includegraphics[scale=0.28,clip=true,trim=0cm 0mm 10mm 0mm]{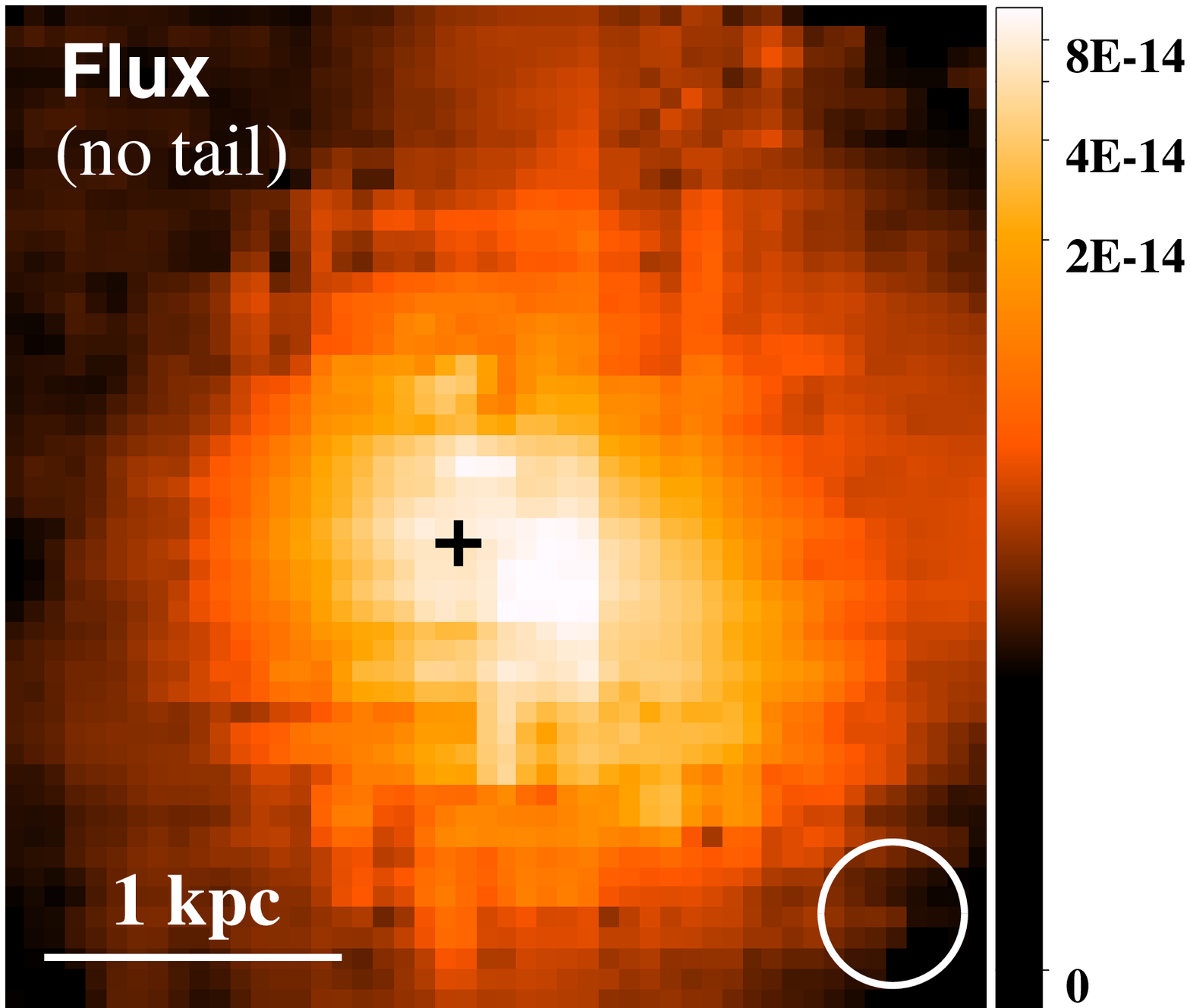}%
\includegraphics[scale=0.28,clip=true,trim=0cm 0mm 18mm 0mm]{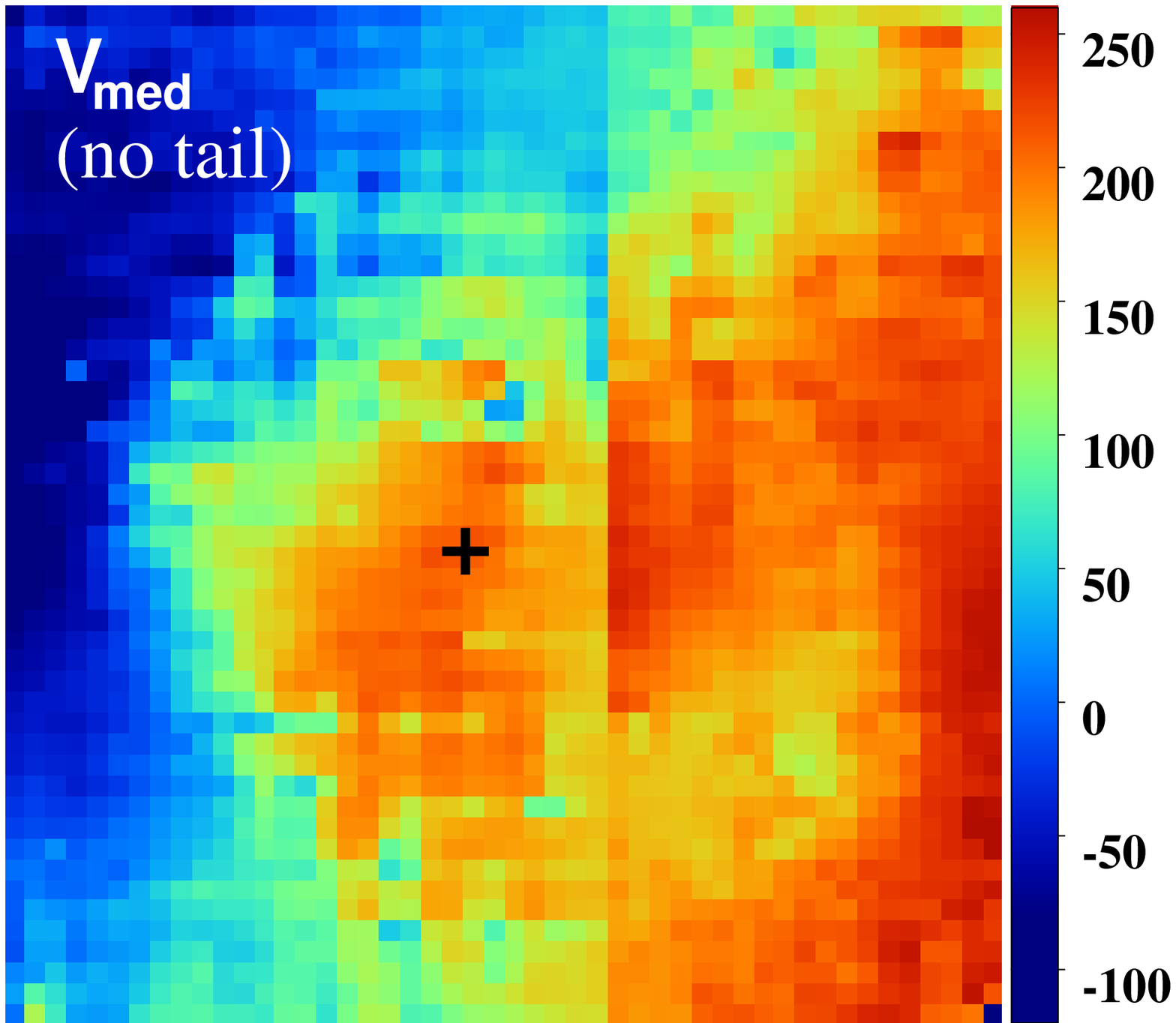}%
\includegraphics[scale=0.28,clip=true,trim=0cm 0mm 25mm 0mm]{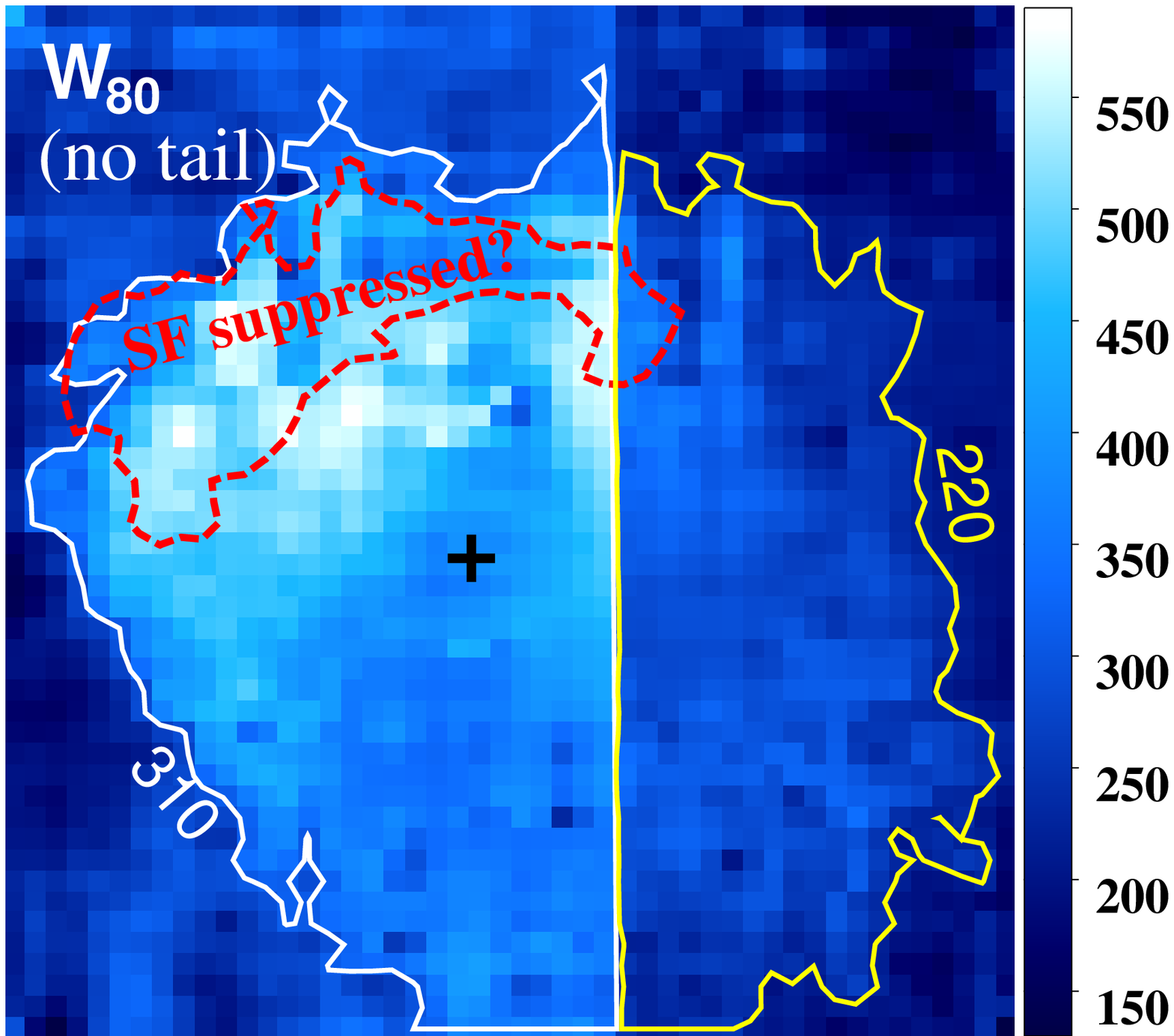}
\includegraphics[scale=0.35,clip=true,trim=4mm 21mm 0mm 0mm]{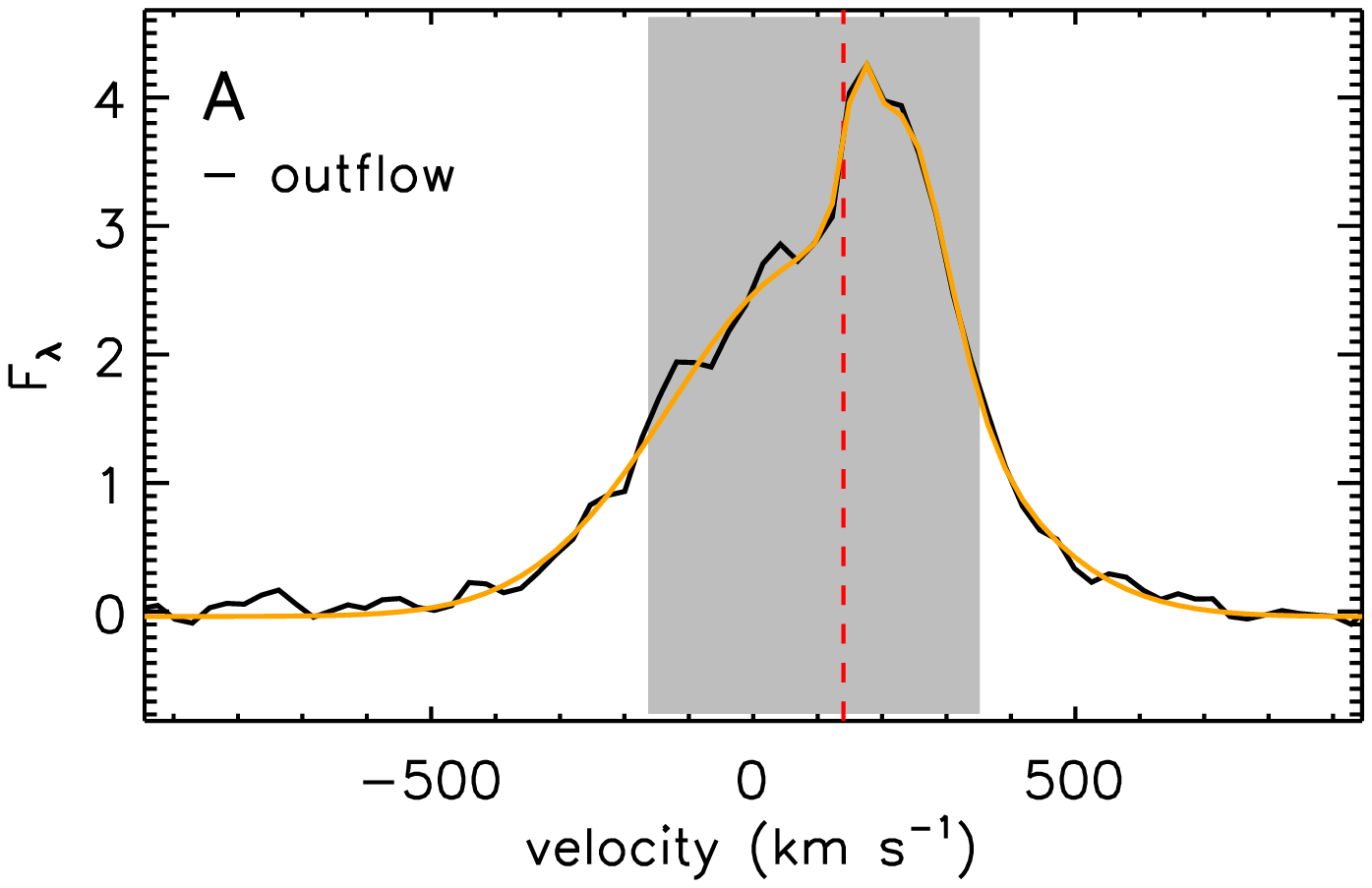}\hspace{1mm}%
\includegraphics[scale=0.35,clip=true,trim=0cm 21mm 0mm 0mm]{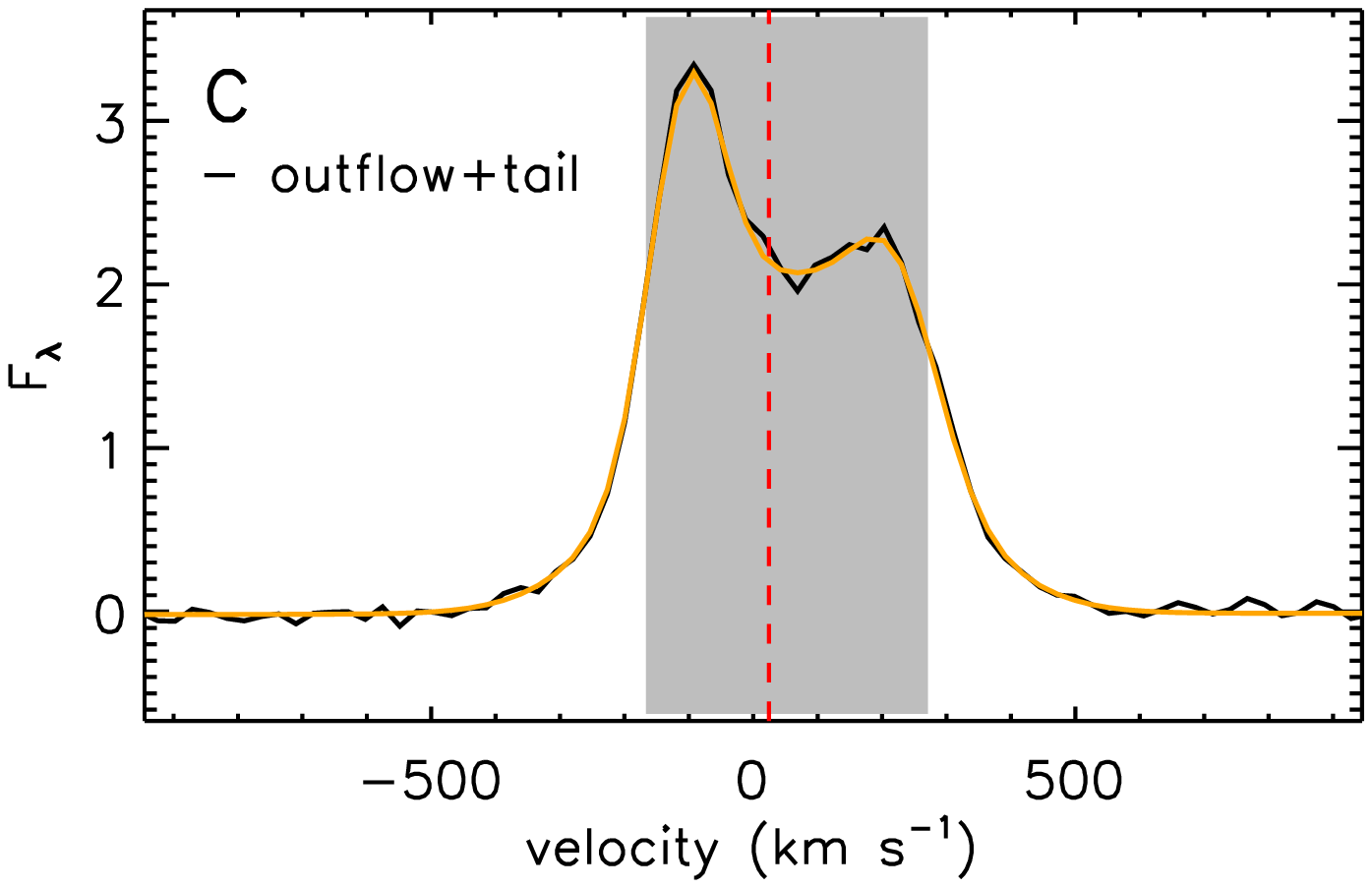}\hspace{1mm}%
\includegraphics[scale=0.35,clip=true,trim=0cm 21mm 0mm 0mm]{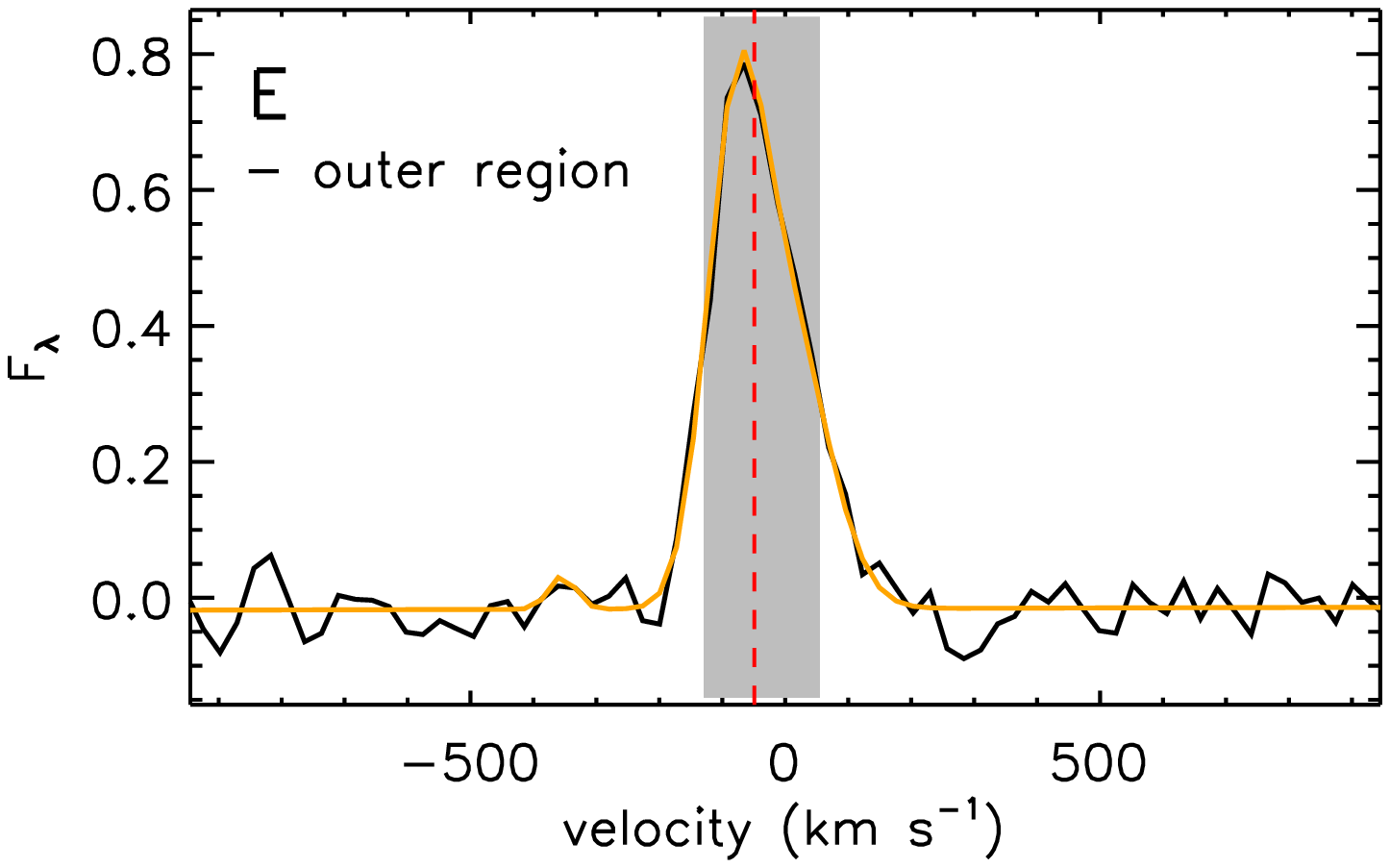}\\
\includegraphics[scale=0.35,clip=false,trim=4mm 5mm 0mm 6mm]{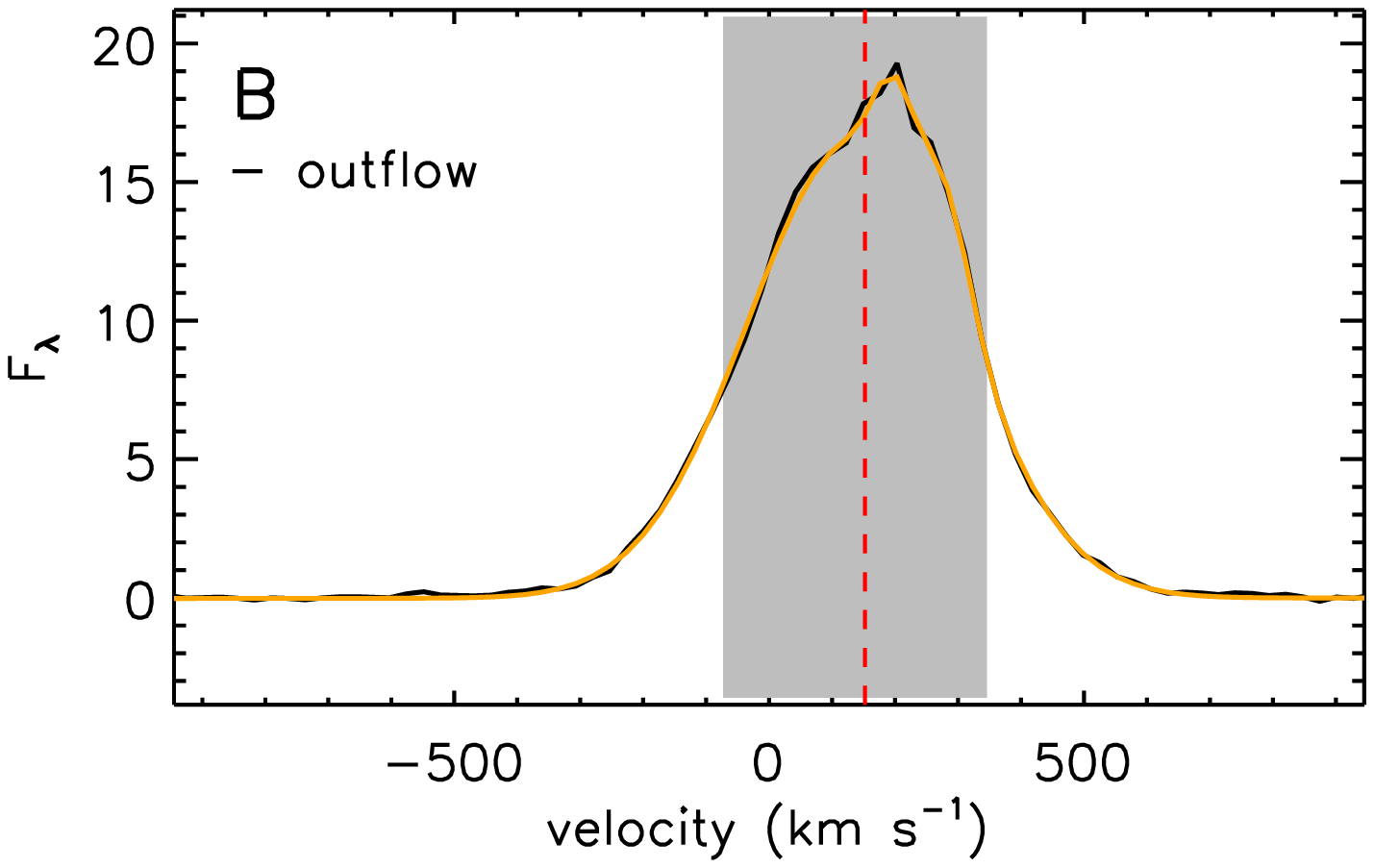}\hspace{1mm}%
\includegraphics[scale=0.35,clip=false,trim=0cm 5mm 0mm 6mm]{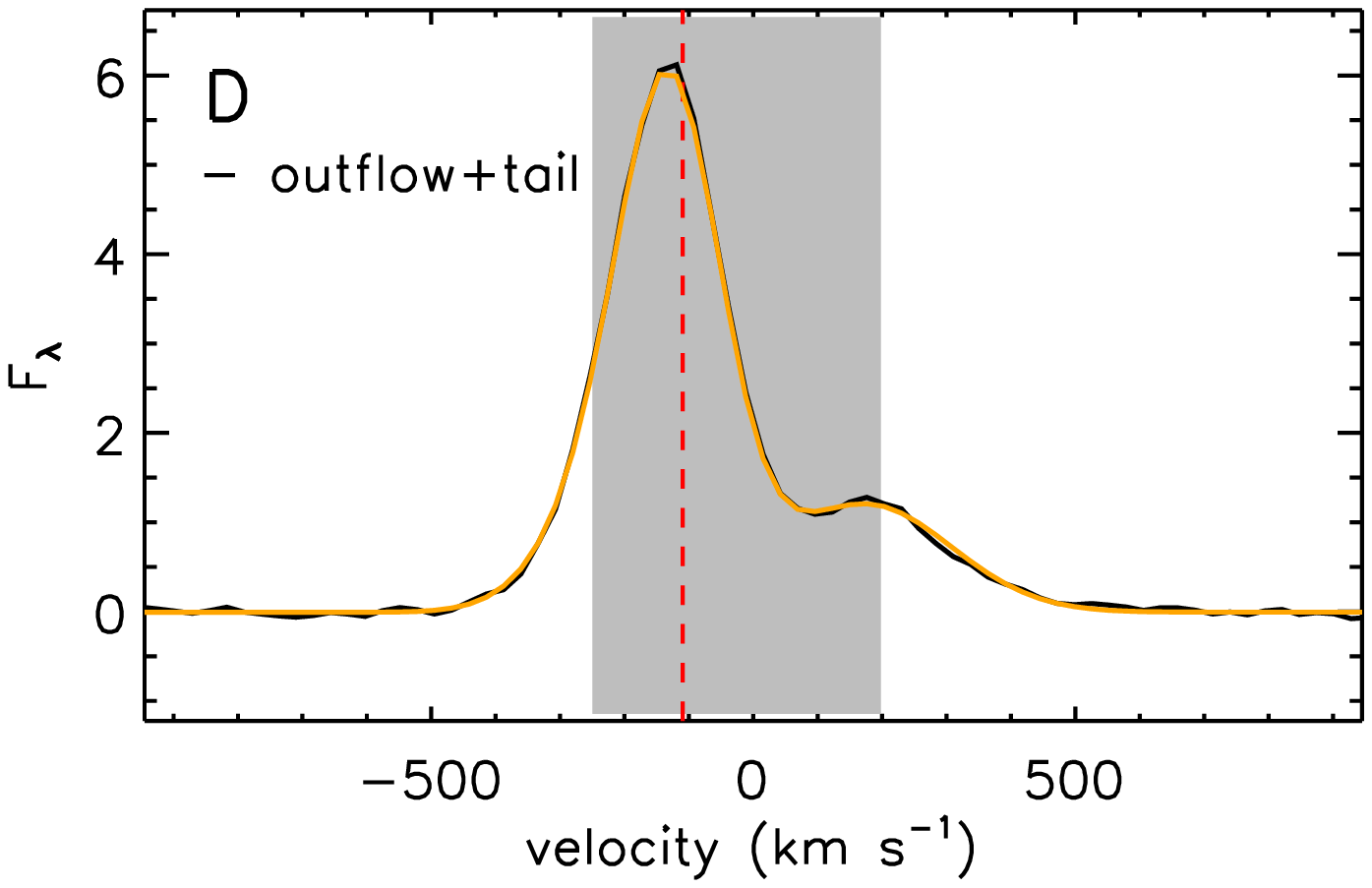}\hspace{1mm}%
\includegraphics[scale=0.35,clip=false,trim=0cm 5mm 0mm 6mm]{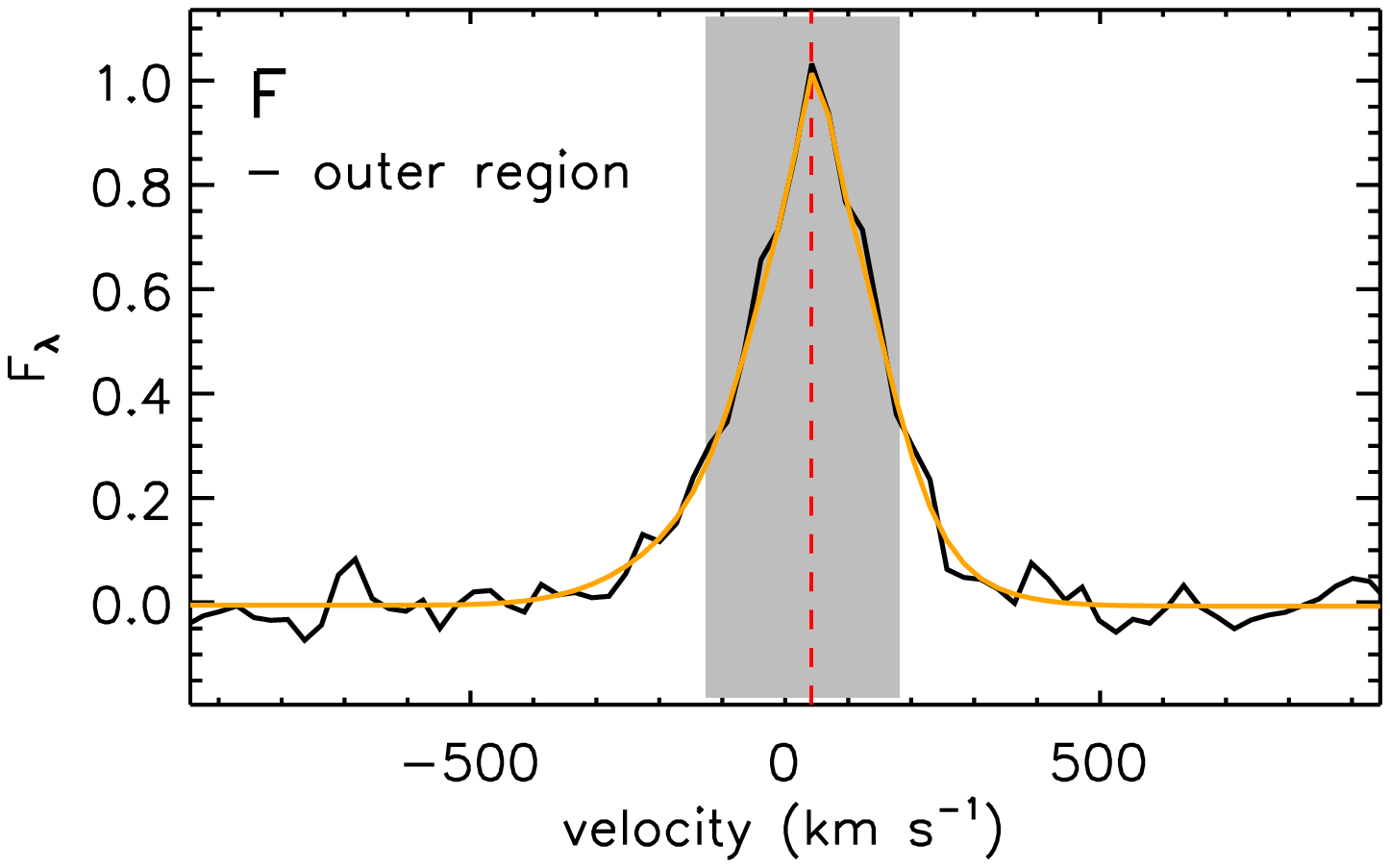}
\caption{Mrk 509. {\bf Top color maps:} 
The \oiii\ surface brightness (left column, in erg s$^{-1}$ cm$^{-2}$ \AA$^{-1}$,
logarithmic scale except for the ``tail''), median velocity 
(middle column, in km s$^{-01}$) and line width ($W_{80}$, right column, 
in km s$^{-01}$) maps. Our 2-Gaussian decomposition of the \oiii\ line allows for isolating
the contribution from the ``tail'' structure (Figure \ref{fig:HST})
and a roughly spherical or elliptical structure likely due to 
an galactic outflow. Overlaid on the bottom right panel is the 220 km s$^{-1}$ 
contour for the ``tail'' region and the 310 km s$^{-1}$ contour for 
the rest of the FoV, which roughly depict the outflow structure.
Red contour crudely marks where the narrow H$\beta$ line is weakest (indicative of
star formation suppression), which spatially coincides with the broadest \oiii\ line
(see discussion in Section \ref{sec:compare}, cf. Figure \ref{fig:s2n}).
Central AGN is marked with a black cross, and the seeing at the observing site 
is shown by open circles. {\bf Bottom spectra:} \oiii\ velocity profiles at
6 representative spatial positions ($F_{\lambda}$ is in units of $10^{-17}$ 
erg s$^{-1}$ cm$^{-2}$ \AA$^{-1}$, within a 0.1\arcsec\ spatial pixel), 
including 2 outflow regions (``A'', ``B''), 
2 outflow plus ``tail'' regions (``C'', ``D''), and 2 outer regions (``E'', ``F''). 
The fitted line is in orange, the median velocity is marked by a red dashed line, 
and the velocity range enclosing $W_{80}$ is a grey box.} 
\label{fig:M509}
\end{figure*}

Mrk 509 is a more complicated system than F04250, and interpreting the \oiii\
nonparametric measurements is nontrivial. 
In particular, the HST FQ508N narrow-band \oiii\ imaging (Figure \ref{fig:HST}, 
\citealt{Fischer13,Fischer15}) detects tidal structures featured by a prominent 
linear tail extending out to a length of 7.8\arcsec\ (5.4 kpc) 
that appears to abruptly bend toward the nucleus from the southwest. \citet{Fischer13}
suggest that the positive velocity of the southwest structure is indicative
of inflowing, rendering the system a minor merger with a dwarf galaxy.
These features are also spatially resolved by our IFU observations, which
becomes more evident after careful PSF subtraction (detailed in Section 
\ref{sec:psf}), and makes characterizing outflow features a challenging 
task (Figure~\ref{fig:M509}, top row).

The \oiii\ line shows unambiguous double-peak profiles 
in the west $\sim$40\% fraction of the FoV. The two peaks are well 
separated wherever the tidal tail dominates the \oiii\ emission: the bluer 
peak always corresponds to a negative velocity (between $-180$ and $-100$ 
km s$^{-1}$), and the redder peak persistently shows a positive velocity 
(70--220 km s$^{-1}$). We therefore perform 2-Gaussian fits to the velocity
profile of the \oiii\ line in every spatial pixel of this region to decompose 
their respective spatial structures. 

This approach results in a clear and physical decomposition demonstrated 
in Figure~\ref{fig:M509}. The intensity map of the Gaussian peaking at
a bluer wavelength reconstructs the morphology of the linear tidal tail precisely, 
whose narrow linewidth (typically $W_{80}\sim200$ km s$^{-1}$) throughout 
the region is also consistent with being an gaseous structure illuminated 
by the central AGN (middle row; see discussion in \citealt{Liu13b} and 
\citealt{Liu09}).
The intensity map of the redder Gaussian reveals a ``southwestern jut''
mentioned in \citet{Fischer13} with little contamination from the
tail (bottom row). 

Here we interpret the ``southwestern jut'' as part of the galactic outflow
(see other possibilities and a discussion of the complex gas kinematics in Section \ref{sec:SIII}).
This is not only because of the wholeness, smoothness and symmetry of \oiii\ flux when it 
is mosaiced with the original data for the rest of the FoV (Figure \ref{fig:M509},
bottom row), but more importantly, because the ratio of \oiii\ to the narrow 
H$\beta$ component is as high as 13 on average (same as that in the center) which
indicates high ionization state caused by AGN emission routinely found in photoionization 
cones (see Section \ref{sec:energetics}; \citealt{Liu13a}), and because the 
mosaiced $W_{80}$ map shows behaviors similar to known outflowing objects (see below).

Broad line widths are indicative of gaseous outflows in IFU data, as reasoned
by \citet{Liu13b}, in which the \oiii-emitting clouds possess a wide local
velocity distribution in addition to the global outflow kinematics, leading
to broad wings of the \oiii\ line profile. 
In absence of the tidal tail, we are left with a structure with elliptical
or roughly round morphology shown by the \oiii\ flux, with relatively high 
$W_{80}$ values distributed in a round-ish region. The situation here is 
similar to the luminous $z\sim0.5$ SDSS quasars observed by \citet{Liu13b}. 
In those objects, high linewidths persist at a roughly constant level 
in the central several kpc ($W_{80}\sim500$--2000 km s$^{-1}$) which often 
decrease to much lower values ($\sim$100--200 km s$^{-1}$) 
in the outer regions (several kpc from the center) beyond the direct 
impact of the outflow. 

In Mrk 509, $W_{80}$ declines rapidly at $\sim310$ km s$^{-1}$ in the 
eastern 60\% of the FoV, and at $\sim220$ km s$^{-1}$ in the region where the 
tidal tail feature is removed through decomposition. 
Because 2-Gaussian fits cannot account for the 
broad base of the \oiii\ line in this region, $W_{80}$ measured there may be
underestimated. Whether or not taking into account this part of the FoV,
we measure the northeastern extent of the outflow (where we are the most 
certain about its outflow characteristic due to the observed suppressed 
narrow H$\beta$ line, see discussion in Section \ref{sec:compare}), and find a radius of 
1.2 kpc (1.7\arcsec) for this roughly spherical (or very wide angle) outflow.

\section{Compare to absorption line analyses}
\label{sec:compare}

\subsection{IRAS F04250--5718}
\subsubsection{Spatial extent} 

The galactocentric distance of the outflow revealed by our Gemini data,
$R\gtrsim2.9$ kpc and $R\gtrsim2.2$ kpc for the two sides of the bi-conical 
outflow, is consistent with the previous UV absorption line
analysis obtained by our group, $R\gtrsim3$ kpc \citet{Edmonds11}. In that
paper, we measured the column density of C {\sc ii} and an upper limit on
the column density of C {\sc ii}$^{\ast}$, and determined the electron density
in the outflow to be $\lesssim$30 cm$^{-3}$. Combined with photoionization 
modeling, we found the above constraint for the outflow radius. The direct
IFU mapping in this work provides the first ever direct test that validates 
our approach to indirectly derive the outflow size, which has been applied on 
nearly 20 objects during the last decade (see \citealt{Arav13} for a review). 

\subsubsection{Gas Kinematics}
\label{sec:F04kinematics}

\citet{Liu13b} proposed a spherical, extinction-free outflow model with a 
constant physical velocity as a representative picture for the quasi-spherical 
outflows seen in their luminous quasars at $z\sim0.5$. For such a simplistic
model, the radial velocity profile remains exactly the same for the entire
system (see Equation 10 therein), rendering the observed linewidth a constant
across the $W_{80}$ map. As discussed in Section \ref{sec:outflow}, the
outflow from IRAS F04250--5718 clearly shows bi-conical structures. The 
opening angle measured from the high $W_{80}$ regions (Figure \ref{fig:F04})
is $\sim70\degr$ for both cones. The spherical model is not directly
applicable to this system, but the calculation in that paper can be utilized 
here.

As in Figure \ref{fig:F04} we do not observe an limb-brightened outflow
\citep[expected for a hollow-cone structure where the walls of the cones 
dominate the emission; e.g.][]{Crenshaw00}, we assume a filled bicone model
equivalent to part of a spherically symmetric outflow, in which the luminosity
density only depends on the spherical radius $r$, and scales as a power 
law, $j(r)\propto r^{-\alpha}$. The very high inclination of the galactic 
disk (see Section \ref{sec:outflow}) allows us to assume that the biconical outflow 
(roughly perpendicular to the galactic disk) to lie approximately in the plane 
of the sky. 

In this model, the maximum linewidth is achieved on the axis of 
the bicone. If we further assume a constant outflow velocity $v_0$, the 
calculation for a spherical model in (\citealt{Liu13b}, Section 4.1, Equations
7, 8 and 9) remains valid. Although the conclusion therein that the radial velocity
profile remains the same at every spatial position is no longer applicable, on 
the axis of the bicone, Equation (10) in that paper only requires a trivial 
modification,
\begin{equation}
I(v_z,R_{\parallel})\propto R_{\parallel}^{1-\alpha}\left[1-\left(\frac{v_z}{v_0 \sin\frac{\theta}{2}}\right)\right]^{(\alpha-3)/2},
\label{eqn:1}
\end{equation}
where $I$ is the \oiii\ flux at a given spatial position and a given wavelength
corresponding to $v_z$, the line-of-sight component of the outflow velocity 
${\bf v_0}$, $R_{\parallel}$ is the projected 2-dimensional radius from the center 
along the outflow axis (in the plane of the sky under our assumption), $\theta$ 
is the opening angle of each cone ($\sim70\degr$), 

In the outer region ($R_{\parallel}\gtrsim1.3$ kpc) of the quasar nebula along
the outflow axis, the observed \oiii\ surface brightness indeed closely follows 
a power-law relation $I\propto R_{\parallel}^{-3.0\pm0.1}$, therefore 
$\alpha=4.0$ as per our best fits. Inserting these parameters into Equation 
\ref{eqn:1}, we find a simple linear relationship between $W_{80}$ and $v_0$,
\begin{equation}
W_{80}=1.374\,v_0\,\sin \frac{\theta}{2}.
\label{eqn:2}
\end{equation}
The observed maximum $W_{80}$ value is found to be 498 and 410 km s$^{-1}$ on 
the axis of the larger and smaller cone, respectively. Due to the limitation 
of our spatial resolution, we estimate the possible range of the opening angle
to be $60\degr\lesssim\theta\lesssim90\degr$, which introduces 
an $\sim$20\% uncertainty in the inferred $v_0$. For $\theta=70\degr$, 
we attain an outflow velocity of $v_0\simeq 520$--630 km s$^{-1}$. 

If the observed $W_{80}$ values are predominantly due to the
projected velocities of the bulk motion of the outflow, the physical velocity
of the outflowing gas must be high. The host galaxy, rotating at a speed of 
$\sim170$ km s$^{-1}$ (measured at 1--2 kpc from the center, see Section 
\ref{sec:outflow}) close to that of the Milky Way ($\sim$220 km s$^{-1}$ at 
the corresponding galactic radius, \citealt{Bhattacharjee14}). 
This rotation speed indicates a host galaxy with comparable 
or slightly smaller mass than the Milky Way (although the rotation we observe 
may not reflect the large-scale dynamical mass of the system, and this 
is only a heuristic comparison).
Considering that the Milky Way has an escape velocity of 
$550.9^{+32.4}_{-22.1}$ km s$^{-1}$ \citep{Kafle14}, the ionized gas traced
by \oiii\ is probably (barely) escaping from the potential well of its host galaxy.
However, we emphasize that in addition to the bulk flow, the local velocity 
distribution of the clouds in the outflow likely also contribute to the observed
linewidth, resulting in a smaller outflow velocity. 

Under the ``Case B'' assumption, the mass of the observed ionized gas
is proportional to the product of H$\beta$ luminosity and electron density
(\citealt{Nesvadba11,Osterbrock06}; though this is actually a lower limit, 
see discussion in \citealt{Liu13b}, Sec. 6). In this object, H$\beta$ follows
the same spatial distribution as that of \oiii\ because \oiii/H$\beta$ is
a constant (Sec. \ref{sec:M509extent}; Fig. \ref{fig:o3hb}).
Assuming that the electron density is distributed as 
$n_{\rm e}=10^3\;(r/0.1\,{\rm kpc})^{-2}$ cm$^{-3}$ (Sec. \ref{sec:M509energy}),
we infer that $\sim$60\% of the warm ionized gas is 
outflowing, therefore likely escaping from the galaxy's potential well.

Absorption line analysis of the high quality HST COS spectrum reveal three 
kinematic components in this system, identified using the strong, unblended 
C {\sc iv} and N {\sc v} doublet lines. Their centroids are found at 
radial velocities of $-38$, $-156$ and $-220$ km s$^{-1}$, respectively 
\citep{Edmonds11}. These absorption outflow components are all at velocities 
smaller than the maximum line-of-sight velocity $v_0\,\sin\,(\theta/2)$ of 
300--360 km s$^{-1}$ (Equation \ref{eqn:2}) given by the bi-cone model built
upon our IFU observations, implying crude consistency. The 
$\gtrsim100$ kms s$^{-1}$ lower velocities in UV absorption are probably 
because these absorbers are far away from the primary biconical structure of 
the ionized gas outflow, indicating non-localized matter distribution and
non-uniform velocity distributions beyond the scope of our simple model.
These poorly understood uncertainties reflect the limitations of current 
knowledge on the physics of outflows, more realistic modeling and further
observational scrutinization in the future are awaited to provide a handle.


\subsection{Mrk 509}

\subsubsection{Spatial extent} 
\label{sec:M509extent}

\begin{figure*}
\centering
\includegraphics[scale=0.85,clip=true,trim=0cm 1mm 30mm 0mm]{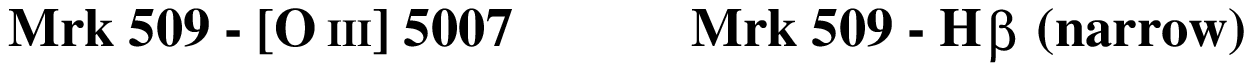}\\
\includegraphics[scale=0.4,clip=true,trim=0cm 0mm 32mm 0mm]{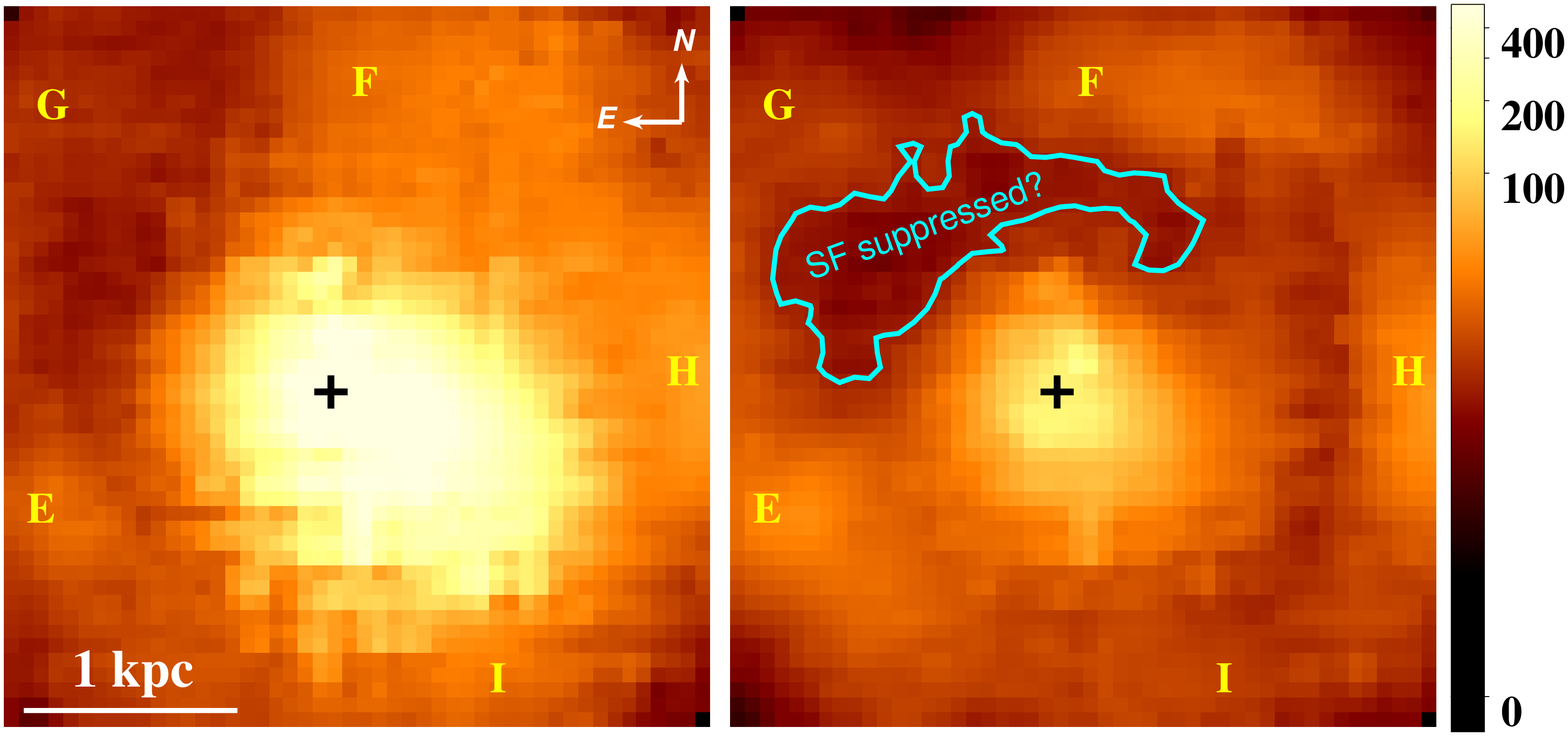}
\vspace*{0.in}
\caption{Maps of signal-to-noise ratios measured at the peak of \oiii\ 5007\AA\ and 
the narrow H$\beta$ emission line profiles. These maps allow for seeing low 
surface brightness features in the outer region more easily than previous flux 
maps (Figure \ref{fig:M509}, left panels). 
At least 5 \oiii-emitting blobs surrounding the center are seen in both panels 
(marked ``E'' to ``I''). Note that the spectra at ``E'' and ``F'' have been shown in 
Figure \ref{fig:M509}. The cyan contour depicts the region of the weakest narrow H$\beta$
emission, which spatially coincides with where $W_{80}$ reaches its maximum values,
indicative of star formation being suppressed by the outflow 
(cf. Figure \ref{fig:M509}, $W_{80}$ maps).
} 
\label{fig:s2n}   
\end{figure*}

In order to compare the distribution of AGN activity to that of 
newly formed stars, we decompose the H$\beta$ line into broad 
(${\rm FWHM}=2700$ km s$^{-1}$) and narrow (${\rm FWHM}=140$--640 km s$^{-1}$)
components using 2-Gaussian fits, so that the emission from the broad
line region and the narrow line region of the AGN is separated (note that
\oiii\ is emitted by the narrow line region only). 
The maps of signal-to-noise ratios at the 
peak of \oiii\ and narrow H$\beta$ lines are shown in Figure \ref{fig:s2n},
where the tidal tail feature has been removed, as described in Section
\ref{sec:outflow}. Faint \oiii-emitting gas illuminated by massive stars 
are characterized by small linewidths (${\rm FWHM}\lesssim200$ km s$^{-1}$). 
Thus, the line intensity in those regions, scaling as the product of the linewidth 
and the peak flux, are thus more suppressed in \oiii\ flux maps than in these 
signal-to-noise ratio maps. As a result, these maps are better for scrutinizing 
low surface brightness features than line intensity maps.

In figure \ref{fig:s2n}, at least 5 \oiii-emitting blobs surrounding the center are seen, 
which are distributed at a galactocentric distance of about 1.6 kpc. All these
features are clearly seen in the narrow H$\beta$ signal-to-noise map at the same 
positions. In galactic environments, hydrogen recombination lines are emitted by 
interstellar gas ionized by young, massive stars with ages $\lesssim10$ Myr 
\citep{Liu13c}, rendering them reliable tracers of the current star formation 
rate \citep{Kennicutt98, Liu11} if dust extinction is carefully corrected for 
spatially-resolved studies (see discussion in \citealt{Calzetti07} and 
\citealt{Liu13d}). Therefore, this spatial correspondence of \oiii\ and
narrow H$\beta$ implies significant contribution or even dominance of massive 
stars in producing the warm ionized gas at this distance. 

Furthermore, the ratio of these two maps reveal that \oiii/H$\beta_{\rm narrow}$ 
is above 10 in the central $\sim1$ kpc and decreases 
to 1--3 at a radius of 1.6 kpc and beyond (see Section \ref{sec:energetics} and 
Figure \ref{fig:o3hb}). In a BPT diagram \citep{Baldwin81}, \oiii/H$\beta_{\rm narrow}>10$ 
implies AGN dominance, while \oiii/H$\beta_{\rm narrow}\sim1$ points to a low-ionization 
nuclear emission-line region (LINER, not applicable to this case) or star formation 
possibly mixed with AGN ionization. The combination of this fact and the
spatial correspondence of \oiii\ and narrow H$\beta$ features strongly indicate that 
the AGN's dominating role in the interstellar radiation is taken over by 
star formation at a distance of $\sim1$--1.5 kpc. We therefore conclude that 
the full extent of predominating AGN ionization likely have been probed, in spite of 
our limited field of view ($3.3\times3.3$ kpc$^2$).

An intriguing phenomenon is the indicative evidence for the outflow 
suppressing star formation: the weakest H$\beta$ (narrow) emission is found 
in an elongated region in the northeast to north direction from the center, 
at a radius a $\sim1$ kpc (delineated with a cyan dashed curve in Figure 
\ref{fig:o3hb}); this region spatially coincides with where the line width 
$W_{80}$ reaches its maximum (Figure \ref{fig:M509}). Similar evidence for 
outflows suppressing star formation activities has been reported in spatially 
resolved spectroscopy at $z=2.4$ \citep{CanoDiaz12}. This further consolidates 
the conclusion in \citet{Liu13b} that high line widths of \oiii\ strongly 
implies outflowing gas.

The multi-wavelength campaign on Mrk 509 has derived a lower limit
of 100--200 pc for the outflow radius in this system \citet{Arav12}. This was
inferred from the fact that the column densities of C {\sc iv} and N {\sc v}
showed negligible variation between 2001 and 2009 despite a large change in
ionizing flux, aided by Monte Carlo simulated light curves that statistically
determined the distance limits. The galactocentric distance of 1.2 kpc that 
we determine from the features of a spherical outflow (Section \ref{sec:outflow})
lies above these indirectly inferred lower limits comfortably.

\subsubsection{Gas Kinematics} 
\label{sec:M509kinematics}

The extinction-free quasi-spherical outflow model proposed in \citet{Liu13b}
is applicable to Mrk 509, because of the round-ish morphology, the
small velocity variation across the median velocity map, and the roughly flat
radial profile of $W_{80}$ within a radius of $\sim1$ kpc. Similar to the
case of IRAS F04250--5718, we measure the outskirt of the ionized gas nebula (before
reaching the star formation dominated region at $R\gtrsim1.3$ kpc), finding
the best-fit power law of \oiii\ surface brightness to be 
$I\propto R_{\parallel}^{-2.9\pm0.1}$. Changing Equation \ref{eqn:1} to its
original form in \citet{Liu13b} (i.e. replacing $R_{\parallel}$ and 
$v_0\,\sin\,(\theta/2)$ with $R$ and $v_0$, respectively), we find for $\alpha=3.9$,
\begin{displaymath}
W_{80}=1.393\,v_0.
\nonumber
\end{displaymath}
To avoid additional uncertainties introduced by removing the linear tail
structure, we only use the west (left) 60\% part of the $W_{80}$ map 
(Figure \ref{fig:M509}) where nonparametric measurements were performed
on the original \oiii\ line profile. The average $W_{80}$ of the outflow
in this region (defined by $W_{80}>310$ km s$^{-1}$) is 
$410\pm72$ km s$^{-1}$, translating to a physical outflow velocity of
$v_0=293\pm51$ km s$^{-1}$, where the standard deviations are reported
as uncertainties. $W_{80}$ reaches it maximum $\sim510$ km s$^{-1}$
in the northeast of the AGN, thus the largest possible outflow velocity
is $v_0\sim370$ km s$^{-1}$.

Since this outflow is quasi-spherical, velocities
from absorption line analyses can be directly compared to our model 
(projection effects are minimized). Nine independent outflow components 
have been identified from the N {\sc v}, Si {\sc iv} and C {\sc iv} 
troughs by \citet{Arav12}. The strongest absorption feature emerges 
at $\sim-320$ km s$^{-1}$, in agreement with our best estimate 
$v_0=293\pm51$ km s$^{-1}$. The highest velocity measured in that
work ($425$ km s$^{-1}$) slightly exceeds the upper bound we derive 
here (370 km s$^{-1}$), but the corresponding component is relatively weak. 
Moreover, as discussed in Section \ref{sec:F04kinematics}, absorption 
line analyses intrinsically have significantly higher spatial resolution 
and are more sensitive to the local velocity distribution of the clouds
residing on our line of sight, which is likely smeared out in the IFU data 
which may only detect the bulk flow. Besides the uncertainty of our simplistic
model, this difference could also be responsible for the small 
discrepancy (several tens of km s$^{-1}$). As a result, these two 
independent and complementary investigations provide consistent results.

\subsection{Electron density}
\label{sec:n_e}

Our IFU spectra include the conventional electron density ($n_{\rm e}$) diagnositic,
[S {\sc ii}] $\lambda\lambda$6717, 6731\AA, facilitating a comparison of $n_{\rm e}$
to our previous absorption line analyses. To do this, we fit the 
[S {\sc ii}] doublet simultaneously using the same line width and velocity 
structure to derive a [S {\sc ii}] 6731\AA/6717\AA\ ratio map. 

The theoretical dependence of this line ratio on $n_{\rm e}$ is then 
calculated using Version 7.1 of {\sc chianti}, an atomic database for spectroscopic 
diagnostics of astrophysical plasmas \citep{Dere97,Landi13}. 
In this calculation, we assume a typical temperature of warm ionized gaseous 
nebulae that produce narrow emission lines, $T=10^4$ K (but note $n_{\rm e}$ 
is only weakly dependent on the assumed temperature as $\propto T^{-1/2}$, 
\citealt{Osterbrock06}). 

As a result, we find $n_{\rm e}=1300\pm140$ cm$^{-3}$ and $1200\pm110$ cm$^{-3}$ 
in the center of IRAS F04250$-$5718 and Mrk 509, respectively.
We limit our $n_{\rm e}$ measurement to locations where the [S {\sc ii}] doublet
has a signal-to-noise ratio of at least 3. At the maximum radii set by the
above requirement, we find $n_{\rm e}<100$ cm$^{-3}$ at 1 kpc away from
the center of Mrk 509, and $n_{\rm e}<200$ cm$^{-3}$ at 2 kpc from the
center of IRAS F04250$-$5718 (these values are upper limits because $n_{\rm e}$ 
already reaches the lower critical density for deexcitation, where the line 
ratio is no longer sensitive to $n_{\rm e}$). In Section \ref{sec:energetics},
$n_{\rm e}$ at the break radius is necessary for deriving the energetics of 
these outflows, which is, unfortunately, not measurable from our data, but 
the adopted value of $10$ cm$^{-3}$ is consistent with these limits.

In the absorption outflow of IRAS F04250$-$5718, a 3$\sigma$ upper limit of 
$n_{\rm e}<30$ cm$^{-3}$ is derived from the constraint on the C {\sc ii}$^*$/C {\sc ii} 
ratio placed by the non-detection of C {\sc ii}$^*$ \citep{Edmonds11}. 
These absorbers are found to be located at $\sim3$ kpc from the center; at this 
distance, we derive a consistent upper limit from [S {\sc ii}] emission lines, 
$n_{\rm e}<200$ cm$^{-3}$.

UV absorption lines allowed \citet{Arav12} to measure electron density for outflow 
components T1 and T2 in Mrk 509, $n_{\rm e}=10^{3.0\mhyphen3.7}$ 
and $10^{2.9\mhyphen3.1}$ cm$^{-3}$, respectively, but the distance for the
absorbers is only constrained to be $>200$ pc. At these distances,
the [S {\sc ii}] emission lines in our data give $n_{\rm e}<900$ cm s$^{-3}$, a result
in broad agreement with \citet{Arav12}, given the $\sim10$\% uncertainty.

\section{Discussion}
\label{sec:discussion}

\subsection{Outflow energetics}
\label{sec:energetics}

In a possibly realistic physical picture, numerous relatively dense 
clouds filling only a fraction of the volume as low as $\lesssim1$\%, 
which are embedded in a rarefied hot wind which might even dominate the 
mass and energy of galactic outflows. At small
galactic radii, the surface of these clouds are ionized by the 
central AGN and emits the observed narrow lines 
(``ionization-bounded''). As the wind carries these clouds to 
larger radii, the decreasing pressure allow them to expand and
gradually become optically-thin, and finally fully ionized at a
certain distance, entering the ``matter-bounded regime''
(cf. discussion in \citealt{Liu13b}). 

This hypothesis of transition is supported by a series of recent observations 
\citep{Liu13a,Liu14,Hainline13,Hainline14}, of which the most
direct is the radial profile of \oiii-to-H$\beta_{\rm narrow}$ 
persisting at $\gtrsim10$ at small radii that starts to decline 
at a break radius of $R_{\rm br}=4$--11 kpc, beyond which 
He {\sc ii}/H$\beta$ increases \citep{Liu13a}. This newly found
phenomenon has enabled \citet{Liu13b} to constrain the estimation 
of the mass flow rate ($\dot{M}$) and kinetic luminosity 
($\dot{E}_{\rm kin}$) of the outflows significantly better
than previous studies. 
In this section, we follow that work to constrain the energetics 
of the two objects.

\begin{figure*}
\centering
\includegraphics[scale=0.6,clip=true,trim=0cm 0mm 0mm 0mm]{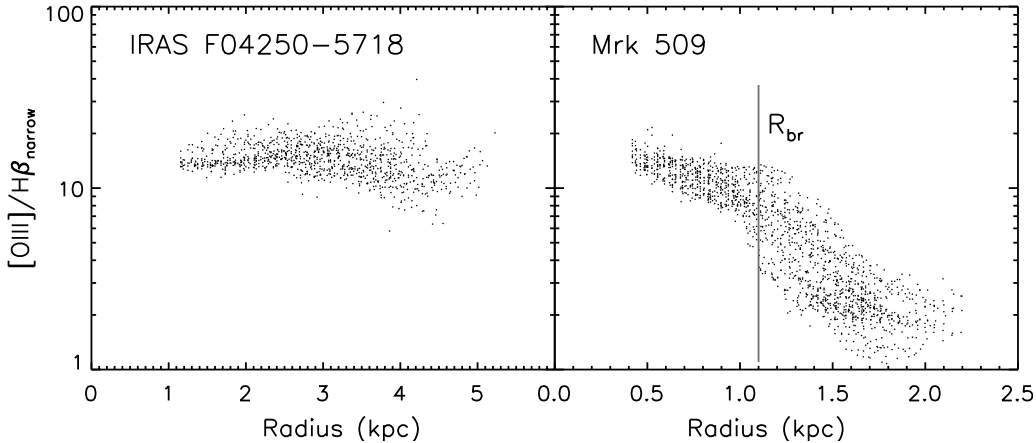}
\vspace*{0.1in}
\caption{The radial profile of the intensity ratio of \oiii5007 to the narrow component of H$\beta$ (obtained
from 2-Gaussian decomposition) for IRAS F04250$-$5718 and Mrk 509.
The ratio persists at a constant level ($\sim$10) in the whole field of view for the former, but starts to 
decline rapidly at a ``break'' radius of 1.1 kpc for the latter (a prevalent phenomenon discovered
in $z\sim0.5$ luminous quasars by \citealt{Liu13a}). This radius (marked with a vertical line) implies a 
transition of the gaseous nebulae from an ionization-bounded regime to a matter-bounded regime, and 
facilitates our calculation of the kinetic luminosity and the mass flow rate of the outflows.} 
\label{fig:o3hb}
\end{figure*}

\subsubsection{Energetics: Mrk 509}
\label{sec:M509energy}

The above \oiii/H$\beta$ behavior is also seen in Mrk 509, 
where the break radius is $R_{\rm br}\simeq1.1$ kpc (Figure \ref{fig:o3hb}).
At this radius, we measure the H$\beta$ surface brightness 
to be $\Sigma_{\rm H\beta}=2.6\times10^{-16}$ erg s$^{-1}$ cm$^{-2}$ arcsec$^{-2}$
after correcting the cosmological dimming effect by a factor 
of $(1+z)^4$. 
This observation facilitates direct application of the
reasoning in \citet{Liu13b}. 

The electron density in outflows is poorly known. In more luminous and energetic objects 
at higher redshifts, \citet{Greene11} find $\sim$100 cm$^{-3}$ in radio-quiet quasar nebulae, 
while \citet{Nesvadba06,Nesvadba08} find a few 100 cm$^{-3}$ in powerful radio galaxies. 
These results are not dissimilar to the density in the sub-kpc neighborhood of our AGNs
(Section \ref{sec:n_e}). The observed \oiii\ surface brightness profile, scaling as $R^{-3}$ 
in both our objects (Section \ref{sec:F04kinematics} and Section \ref{sec:M509kinematics}), is a 
natural consequence of $n_{\rm e}\propto r^{-2}$ \citep{Liu13a}. 
In both objects, the [S {\sc ii}] emission lines give $n_{\rm e}\sim1000$ cm$^{-3}$ at a 
galactocentric radius of 0.1 kpc; it will fall to $\sim10$ cm$^{-3}$ at the break radius
(1.1 kpc for Mrk 509). Although we can only measure an upper limit at $\sim1$ kpc 
($n_e<100$ cm$^{-3}$; Section \ref{sec:n_e}), this extrapolation leads to $n_{\rm e}$ 
similar to \citet{Rupke13}; those authors assume
$n_{\rm e}=10$ cm$^{-3}$ based on measurements of the supperbubble in NGC 3079
($n_{\rm e}=5$--100 cm$^{-3}$; \citealt{Veilleux94,Cecil01}). The above considerations
lead us to adopt $n_e=10$ cm$^{-3}$ for the following calculations.

We find the H$\beta$ line emission to fall off as $\Sigma_{\rm H\beta}\propto R^{-2.1}$
in this object, changing the numerical factor 1.44 in Equation 18 in \citealt{Liu13b} 
to 1.94. Using Equation 22 therein, we estimate the mass flow rate to be
\begin{eqnarray}
\frac{\dot{M}}{\rm 5.2\,M_{\odot}\,{\rm yr}^{-1}}=
\frac{\Sigma_{\rm H\beta}(R_{\rm br})}{2.6\times 10^{-16}\,{\rm erg\,s^{-1}\,cm^{-2}\,arcsec^{-2}}} \nonumber \\
\times\frac{R_{\rm br}}{1.1\,{\rm kpc}}\;\frac{v_0}{293\,{\rm km\,s^{-1}}}
\left(\frac{n_e}{10\,{\rm cm}^{-3}}\right)^{-1},
\label{eqn:mass}
\end{eqnarray}
which is normalized by our derived outflow velocity, 293 km s$^{-1}$ (Section \ref{sec:compare}). 
As a consequence, the kinetic luminosity of the outflow, $\dot{E}_{\rm kin}=\dot{M}v_0^2/2$, is given by
\begin{eqnarray}
\frac{\dot{E}_{\rm kin}}{1.4\times 10^{41}\,{\rm erg\,s^{-1}}}=
\frac{\dot{M}}{\rm 5.2\,M_\odot\,yr^{-1}} \left(\frac{v_0}{293\,{\rm km\,s^{-1}}}\right)^2.
\label{eqn:energy}
\end{eqnarray}
The momentum flow rate is $\dot{P}=\dot{M}\,v=9.6\times10^{33}$ dyne, or 
$\log\,(c\,\dot{P}/L_{\odot})=10.9$, similar to that of the outflows driven by low-redshift
major mergers (\citet{Rupke13} find a range of 10.0--12.8).

The uncertainty of these results is largely caused by the poorly constrained $n_{\rm e}$.
$\dot{M}$ and $\dot{E}_{\rm kin}$ for other electron densities can be easily calculated using
the $n_{\rm e}^{-1}$ dependence (Equations \ref{eqn:mass} and \ref{eqn:energy}).
The conservative constraint that \oiii-emitting clouds only fill a fraction of
the volume results in a lower limit of $n_{\rm e}\sim0.1$ cm$^{-3}$ (Equation 21 in \citealt{Liu13b}). 
Along with the measured upper limit, $n_{\rm e}<100$ cm$^{-3}$ (Section \ref{sec:n_e}), we
find firm lower limits using Equations \ref{eqn:mass} and \ref{eqn:energy},
\begin{displaymath}
0.5\;{\rm M_{\odot}\;yr^{-1}}<\dot{M}< 520\;{\rm M_{\odot}\;yr^{-1}},
\end{displaymath}
\begin{displaymath}
1.4\times10^{40}\;{\rm erg\;s^{-1}}<\dot{E}_{\rm kin}< 1.4\times10^{43}\;{\rm erg\;s^{-1}}.
\end{displaymath}

Hence, we conclude that Mrk 509 is driving an outflow with $\dot{M}\sim 5$ M$_{\odot}$ yr$^{-1}$
and $\dot{E}\sim 1\times10^{41}$ erg s$^{-1}$, with an upper (lower) limit of 100 (10) times 
larger (smaller) for both quantities. The ratio of the momentum flux of the outflow to the AGN 
radiation is $\dot{P}/(L_{\rm bol}/c)\sim0.3$. The outflow kinetic luminosity is only $\sim0.01$\% 
of the bolometric luminosity of Mrk 509, and $\sim0.002$\% of its Eddington luminosity 
($L_{\rm bol}=1\times10^{45}$ erg s$^{-1}$, $L_{\rm bol}/L_{\rm Edd}\simeq0.16$; \citealt{Kaspi00}). 
Since theoretical modeling predicts that significant feedback requires $\gtrsim 0.5$--5\% of the AGN’s 
luminosity to be converted to the mechanical energy of the outflow \citep[e.g.][]{DiMatteo05,Hopkins10},
the outflow in Mrk 509 does not produce feedback strongly affecting the evolution of the host galaxy.

On the absorption line front, \citet{Arav12} find an average hydrogen column density of 
$N_{\rm H}=10^{19.8}$ cm$^{-2}$ and a velocity of $\sim300$ km s$^{-1}$ in the outflow of Mrk 509.
Adopting a solid angle of 4$\pi$ for the quasi-sphere geometry and an outflow radius of 1.2 kpc
obtained in this work, we find $\dot{M}=3.3$ M$_{\odot}$ yr$^{-1}$ (Equation 7, \citealt{Edmonds11}), 
and $\dot{E_{\rm kin}}=\dot{M}\,v^2/2=9.3\times10^{40}$ erg s$^{-1}$, in agreement with the above
determination within a factor of 1.6.

\subsubsection{Energetics: IRAS F04250$-$5718}
\label{sec:F04energy}

The uniform \oiii-to-H$\beta_{\rm narrow}$ ratio $\gtrsim10$ across the 
whole map implies a break radius beyond the largest radius reached by
our field of view ($R_{\rm br}\gtrsim 5.5$ kpc, see Figure \ref{fig:o3hb}).
However, the largest break radius that has ever been observed is 11 kpc,
seen in SDSS J032144.11$+$001638.2 at $z=0.643$ \citep{Liu13a}, one of 
the most luminous quasars at that redshift, with a bolometric luminosity 
of $\gtrsim10^{46.5}$ erg s$^{-1}$. IRAS F04250$-$5718 is expected to
have $R_{\rm br}\lesssim10$ kpc due to its lower luminosity 
($L_{\rm bol}\sim9\times10^{45}$ erg s$^{-1}$, \citealt{Edmonds11}).

At the 5.5 kpc radius, the \oiii\ surface brightness is measured to be 
$1.3\times10^{-15}$ erg s$^{-1}$ cm$^{-2}$ arcsec$^{-2}$ after correcting 
for the cosmological dimming effect, thus the surface brightness profile 
$\Sigma_{\rm [O\,\scriptscriptstyle{III}]}\propto R^{-3.0\pm0.1}$ 
(Section \ref{sec:F04kinematics}) predicts 
$\Sigma_{\rm [O\,\scriptscriptstyle{III}]}(11\,{\rm kpc})=1.7\times10^{-16}$
erg s$^{-1}$ cm$^{-2}$ arcsec$^{-2}$, and thus
$\Sigma_{\rm H\beta}(11\,{\rm kpc})\gtrsim 
0.1\;\Sigma_{\rm [O\,\scriptscriptstyle{III}]} 
=1.7\times10^{-17}$ erg s$^{-1}$ cm$^{-2}$ arcsec$^{-2}$.

Compared to a spherically symmetric geometry, the biconical configuration
of the outflow introduces a correction factor of
$1-\cos\,(\theta/2)$, where $\theta\simeq70\degr$ is the opening angle
of each cone (Section \ref{sec:outflow}).
Now that both $\Sigma_{\rm [O\,\scriptscriptstyle{III}]}$ and 
$\Sigma_{\rm H\beta}$ fall off as $R^{-3}$, the numerical
factor 1.44 in Equation (18) in \citet{Liu13b} increases to 1.57.  
For an outflow velocity of 520 km s$^{-1}$ (the lower bound, Section
\ref{sec:F04kinematics}), assuming $R_{\rm br}=11$ kpc and $n_e=10$ cm$^{-3}$, 
Equation \ref{eqn:mass} gives 
\begin{displaymath}
\dot{M}> 1.3\;{\rm M_{\odot}\;yr^{-1}},
\end{displaymath}
\begin{displaymath}
\dot{P}> 4.3\times10^{33}\;{\rm dyne},\;\;\;
\dot{E}_{\rm kin}> 1.2\times10^{41}\;{\rm erg\;s^{-1}}.
\label{eqn:F04MassEnergy}
\end{displaymath}
As discussed in Section \ref{sec:M509energy}, compared to Mrk 509, this object 
has similar $n_{\rm e}$ in the center and similar $n_{\rm e}$ radial profile, 
but its break radius is larger by a factor of a few or more, and $n_{\rm e}$
at the break radius is likely lower than 10 cm$^{-3}$. However, only an upper 
limit, $n_{\rm e}<200$ cm s$^{-1}$, can be derived from our data, after all.
Here we adopt $n_e=10$ cm$^{-3}$ for consistency (which pushes the above 
results to their lower limits, see below).

Since the actual electron density is likely lower, and the smallest 
possible $\Sigma_{\rm H\beta}$ is used here; although $R_{\rm br}$ may have 
been overestimate by a factor of $\leqslant2$, the overly large break radius 
leads to an underestimate of the steep declining $\Sigma_{\rm H\beta}$ 
(by a factor of $\leqslant8$) that itself cannot balance out. For the above 
reasons, Equation \ref{eqn:F04MassEnergy} gives lower limits of the outflow energetics.

On the other hand, adopting $n_{\rm e}=0.1$ cm$^{-3}$, $R_{\rm br}=5.5$ kpc 
and $v_0=630$ km s$^{-1}$ (upper bound) sets conservative upper limits
\begin{displaymath}
\dot{M}<690\;{\rm M_{\odot}\;yr^{-1}},
\end{displaymath}
\begin{displaymath}
\dot{P}< 2.7\times10^{36}\;{\rm dyne},\;\;\;
\dot{E}_{\rm kin}<8.6\times10^{43}\;{\rm erg\;s^{-1}}.
\end{displaymath}

The ratio of the momentum flux of the outflow to the AGN radiation 
is only loosely contrained to $\dot{P}/(L_{\rm bol}/c)=0.01$--9. 
The kinetic luminosity of the outflow in this system is insignificant,
being only $\gtrsim0.002$\% of its bolometric luminosity ($\sim9\times10^{45}$ 
erg s$^{-1}$, \citealt{Edmonds11}). Therefore, the outflow in 
IRAS F04250$-$5718 is not a significant feedback agent, similar to
the case of Mrk 509.

\citet{Edmonds11} find a velocity of 220 km s$^{-1}$ and a column density 
lower limit of $N_{\rm H}>10^{19.55}$ cm$^{-2}$ in the absorption outflow of this object.
Adopting an outflow radius of 3 kpc, a solid angle of $\Omega/4\pi=0.2$ 
derived from our IFU data and deprojecting their velocities using $\theta=70\degr$ 
(the outflow lies approximately in the plane of the sky, Section \ref{sec:F04energy}), their results lead to
$\dot{M}\sim1.2$ M$_{\odot}$ yr$^{-1}$, and $\dot{E}_{\rm kin}\sim5.5\times10^{40}$ erg s$^{-1}$,
in agreement with the above determination within a factor of $\sim2$.



\subsection{More on Mrk 509's structure}
\label{sec:SIII}

Recent long-slit \citep{Fischer13} and IFU spectroscopy \citep{Fischer15} demonstrate 
that Mrk 509 is likely a minor merger system, where gas is inflowing and fueling the
central AGN through the linear tidal tail and the ``southwestern jut'' (cf. Figure \ref{fig:HST}).
In this section, we discuss in detail the dynamical structure of this system by 
combining and comparing our data to previous observations.
  
The host galaxy is nearly face-on ($i=36\fdg4$ as per the HyperLeda 
database\footnote{http://leda.univ-lyon1.fr}, \citealt{Paturel03}; 
\citealt{Yaqoob03} find $i=41\degr$), and the linear tidal tail is likely 
located in front of the disk, as otherwise it would have been severely 
extincted by dust and become invisible. (To test this judgement, we
isolate the tidal tail's emission in the H$\alpha$ and H$\beta$ lines
using 3-Gaussian fits accounting for the broad base and the double-peak
structure, and find the average H$\alpha$/H$\beta$ ratio on the tail to 
be $2.8\pm0.5$. Under ``Case B'' assumption, the intrinsic line ratio 
is 2.7--3.0 for a wide range of temperature and electron density
($T=5000$--30000 K, $n_{\rm e}=10^2$--$10^6$ cm$^{-3}$). Hence, the 
observed line ratio is consistent with negligible dust extinction, 
confirming our conclusion that the galactic disk is behind the tidal tail.)
Therefore, in case the ``southwestern jut''
is physically connected to the tail and serves as a tunnel for inflowing gas, 
we expect to see velocity smoothly transition from $\sim-100$ km s$^{-1}$ 
to zero then to positive velocities, along the route of the tail, then the
``jut'', finally the AGN neighborhood. 

Nevertheless, in our data we find that
the ``jut'' and the long linear tail, although appear to be connected, 
have distinct velocities everywhere (different by 200--350 km s$^{-1}$). 
Furthermore, the \oiii\ intensity map of the tidal tail (Figure \ref{fig:M509}),
middle row) shows that it likely switches direction sharply (southward to eastward)
to approach the nucleus before it reaches the main body of the ``jut'', at
only $\sim0.4$\arcsec\ (R.A. distance) south of the AGN. This phenomenon is
confirmed by IFU maps with higher spatial resolution \citet{Fischer15}.
Hence, we conclude that the linear tail and the ``jut'' are physically unrelated, 
but are superposed by projection effects; the gas in the linear tail is likely 
indeed inflowing \citep{Fischer13}, but it may use its own route 
instead of inflowing through the ``southwestern jut''.

\begin{figure}
\centering
\includegraphics[scale=0.38,clip=true,trim=0cm 0mm 22mm 0mm]{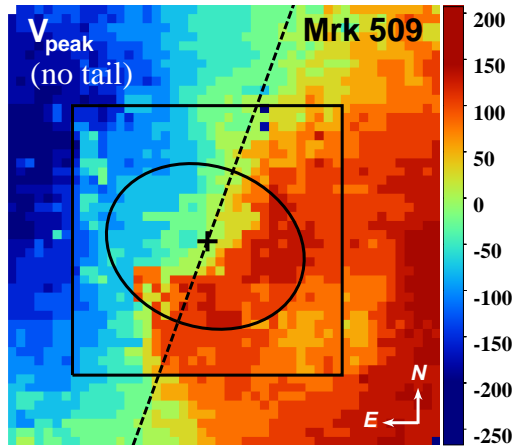}
\caption{\oiii 5007\AA\ emission line's peak velocity map of Mrk 509 (in km s$^{-1}$;
the velocity at the AGN position is set to be zero). 
In contrast to the median velocity map (more sensitive to the broad base of the line 
profile, Figure \ref{fig:M509}) that shows a roundish structure in the center likely
caused by an outflow, this map exhibits a well defined 
large-scale rotation pattern. \citet{Fischer15} find a similar rotating pattern
in the [S {\sc iii}] 0.95 \micron\ emission line, with an axis (dashed 
line) in consistency with this map; those authors attribute this motion to a
rotating disk $\sim2$\arcsec\ in diameter (schematically shown as an ellipse) 
in the center of their $3\arcsec\times3\arcsec$ FoV (box), but we see the 
rotation taking place on a larger scale, co-existing with a quasi-spherical or
wide-angle outflow. However, the uncertainties that the rotating gas introduces
into our outflow modeling are insignificant (see Section \ref{sec:SIII}).} 
\label{fig:pkvel}
\end{figure}

However, with the linear tidal tail removed using 2-Gaussian fits, interpreting 
the remaining structures (including the ``jut'') is still nontrivial. 
\citet{Fischer15} performed IFU spectroscopy assisted by adaptive optics in 
the [S {\sc iii}] 0.95 \micron\ emission line. In their data with higher
($\sim$0.1\arcsec) angular resolution, they observe a $\sim$2\arcsec\ elliptical
region (schematically shown by a ellipse in Figure \ref{fig:pkvel}) in their smaller 
FoV (3\arcsec$\times$3\arcsec). This region shows a well defined 
velocity gradient (upper left panel of Figure 4 in their paper), allowing them
to fit the velocity field to a rotating disk model. 

However, this rotation pattern
is not seen in our median velocity maps (Figure \ref{fig:M509}): we do observe
negative velocities close to the east edge of the FoV, but we see a roughly constant 
velocity across a large round-ish region when the tail is removed, which is deemed as
indicative evidence for a quasi-spherical outflow in Section \ref{sec:M509outflow}.

This seeming discrepancy is, in fact, due to the higher signal-to-noise ratios of 
our data. In fact, if we determine the velocity field from the peak wavelength of the 
\oiii\ line instead of calculating the median velocity, the pattern seen by \citet{Fischer15}
is reproduced with accuracy. As shown in Figure \ref{fig:pkvel}, the rotation axis
of their best-fit model (dashed line) is consistent with our zero-velocity spatial 
pixels (here we force the velocity at the center to be zero for comparison purposes).
Obviously the rotation pattern is actually not restricted within a circumnuclear disk 
($\sim 0.7$ kpc in radius) as \citet{Fischer15} find, but exists on a scale as wide 
as $3.3\times3.3$ kpc$^2$ (our whole FoV) or larger. This finding may alleviate the 
issue of the rather high rotation speed those authors derived ($\sim500$ km s$^{-1}$), 
if their model fits are performed on an area larger by a factor of $\sim5$.

In \citet{Fischer15}, [S {\sc iii}] is detected with ${\rm S/N}<3$ in $\sim40$\% area 
of the FoV (roughly between the ellipse and the box in Figure \ref{fig:pkvel}), where
we detect \oiii5007\AA\ line with ${\rm S/N}>15$ in every spaxel (Figure \ref{fig:s2n}). 
In low S/N data, the weak broad base of an emission line, which 
may contain outflow signatures, can be easily submerged into noise, so that a single
Gaussian is sufficient for line fitting (this effect has 
been simulated and discussed in \citet[][Section 3.4]{Liu13b}.
This difference explains why in majority of their spaxels [S {\sc iii}] can be fitted 
with a single Gaussian profile, which only happens in a few percent of the spaxels in 
our \oiii\ analysis. It also results in our velocity map different from the one obtained
by \citet{Fischer15}, as median velocity is more sensitive to the broad base than
fits to lower S/N data, and thus likely traces outflow kinematics better.

Meanwhile, \citet{Fischer15} also detect a second Gaussian component northeast to the 
AGN, which they suspect is an inflow or outflow. We now obtain supportive evidence for 
an outflow --- this region spatially coincides with where \oiii\ is broadest and 
where the narrow H$\beta$ is weakest (Figures \ref{fig:M509} and \ref{fig:o3hb}), 
indicating star formation being suppressed by an outflow. This northeast outflow region 
shows an opening angle of $\sim90\degr$ in the data of \citet{Fischer15}, whereas
our data points to a significantly wider angle or even a spherical morphology 
(Section \ref{sec:M509outflow}).

The above considerations lead us to conclude that a wide-angle outflow and a large-scale
rotating gas likely both contribute to the kinematics (beside the inflowing tidal tail
in the west). Decomposing the two components is difficult, either spatially or 
spectroscopically, but we reckon that the contribution from the rotating gas is 
not significant. Empirically, rotating gas in the host galaxy produces emission lines 
narrower than those from outflows ($W_{80}\lesssim200$ km s$^{-1}$ vs. $W_{80}\gtrsim300$--400 
km s$^{-1}$). Therefore, in general, $W_{80}$ directly measured from the observed
line profile is not significantly different from the line width of the outflow only
(unless the outflow is remarkably fainter than the rotating gas, in which case the 
measured $W_{80}$ provides a lower limit). 
This assertion is supported by the existing data: \citet{Fischer15}
separate out the outflow contribution using two-Gaussian fits and finding the outflow
component in the northeast region to have $W_{80}=500$--600 km s$^{-1}$, in 
agreement with our direct measurement from the observed \oiii\ line (Figure \ref{fig:M509}).
Hence, the line width used for calculating the outflow velocity in Section \ref{sec:M509outflow} 
is confirmed by an independent observation, and the contamination from the rotating galactic
gas is trivial.

Thus far, we have assumed a spherical geometry for the outflow in Mrk 509. As discussed
above, the northeast quadrant is where the existence of an outflow is supported by the
strongest evidence. Since the firm lower limit of its opening angle is $\sim90\degr$
as given by \citep{Fischer15}, assuming such a biconical geometry with its axis close to the plane
of the sky leads to a velocity larger by $\sim\sqrt{2}$ and an additional filling factor 
of $\sim0.3$, and thus a mass flow rate and a kinetic luminosity not dissimilar 
(within a factor of $\sim2$) to the results of assuming a spherical symmetry.

\section{Summary}
\label{sec:summary}

This work is the first effort to bridge integral field spectroscopy and absorption
line analysis investigations on the ionized phase of AGN outflows, the two 
independent and complementary approaches to determine outflow properties. We present 
Gemini IFU observations of the ionized gas nebulae around two low-luminosity quasars, 
Mrk 509 and IRAS F04250$-$5718.
The outflow distances from the central AGNs have been indirectly determined from UV 
absorption line analyses, which are directly tested by our IFU observations in this 
work.

We demonstrate that IRAS F04250$-$5718 shows a clear case where a biconical outflow
extending out to at least 2.9 kpc from the center, in agreement with the result of
absorption line analysis ($R\gtrsim3$ kpc). The opening angle of each cone 
is $\sim70\degr$, and the outflow axis is $\sim20\degr$ away from the axis
of the galactic disk rotating at a speed of $\sim170$ km s$^{-1}$. The outflowing gas, 
with a derived physical velocity of $580\pm80$ km s$^{-1}$, is probably (barely) 
escaping from the potential well of its host galaxy. We obtain lower limits of the 
mass flow rate, $\dot{M}>1.3$ M$_{\odot}$ yr$^{-1}$, the momentum flux rate
$\dot{P}> 4.3\times10^{33}$ dyne, and the kinetic luminosity 
$\dot{E}_{\rm kin}>1.2\times10^{41}$ erg s$^{-1}$.

Mrk 509 is a more complicated system especially because of a linear tidal tail.
After removing the tidal tail spectroscopically, we find a quasi-spherical outflow with
a radius of $\sim1.2$ kpc, also consistent with pre-existing absorption line analysis 
(a lower limit of 100--200 pc). The spatial coincidence of the highest \oiii\ line widths and 
the weakest narrow H$\beta$ emission is indicative evidence for the outflow suppressing 
the formation of stars. Using the derived outflow velocity of $290\pm50$ km s$^{-1}$, we
estimate a mass flow rate of $\dot{M}\sim5.2$ M$_{\odot}$ yr$^{-1}$, a momentum flux rate of
$\dot{P}\sim9.6\times10^{33}$ dyne, and a kinetic luminosity of 
$\dot{E}_{\rm kin}\sim1.4\times10^{41}$ erg s$^{-1}$.

Adopting the outflow radii and geometric parameters measured from IFU, absorption
line analyses would yield mass flow rates and kinetic luminosities that are in agreement with 
our IFU determination within a factor of $\sim2$. Measuring the outflow properties is 
challenging, and the multiple difficulties lead to relatively large uncertainties. 
However, we emphasize that the lower and upper limits we derived in this work spanning wide 
ranges are rather (maybe even overly) conservative, and should be understood as firm limits 
instead of limits with statistical meanings (e.g. $3\sigma$ limits). The measured values of 
the outflow rates and energetics are reasonably reliable, and the above agreement between 
IFU and absorption line analyses is a justified validation.

The [S {\sc ii}] $\lambda\lambda$6717,6730 \AA\ doublet allows us to derive electron 
densities in the central 1-2 kpc of both objects. The derived electron densities are 
in agreement with the previously inferred values from their absorption outflows 
(Mrk 509: $n_{\rm e}\sim10^3$ cm$^{-3}$ at $\sim$100 pc from the center; 
IRAS F04250$-$5718: $n_{\rm e}<200$ cm$^{-3}$ at $\gtrsim2$ kpc).

We conclude that both the spatial locations and the kinematics of these outflows
determined from IFU spectroscopy of emission lines are consistent with those
derived from previous UV absorption line analyses. Theoretical modeling predicts 
that significant feedback requires $\sim0.5$--5\% of the AGN's Eddington 
luminosity to be converted to the mechanical energy of the outflow.
Feedback in these systems is taking place on kpc or galaxy-wide scales, but is 
inadequate for regulating the evolution of their galaxy hosts effectively 
(kinetic luminosity is only $\gtrsim0.002$--0.01\% of the bolometric luminosity), 
unless it used to be significantly more energetic in the past.

\acknowledgments

We thank Gerard A. Kriss and Steven B. Kraemer for helpful discussions.
We are grateful to the anonymous referee for a careful reading of the manuscript.
G.L. and N.A. acknowledge support from NSF grant AST 1413319 as well
as NASA STScI grants GO 11686 and GO 12022.
D.S.N.R. was supported by a Cottrell College Science
Award from the Research Corporation for Science Advancement.

Based on observations obtained at the Gemini Observatory (acquired through the Gemini 
Science Archive and processed using the Gemini IRAF package, Program ID: GS-2013B-Q-84), 
which is operated by the Association of Universities for Research in Astronomy, Inc., 
under a cooperative agreement with the NSF on behalf of the Gemini partnership: the 
National Science Foundation (United States), the National Research Council (Canada), 
CONICYT (Chile), the Australian Research Council (Australia), Minist\'{e}rio da Ci\^{e}ncia, 
Tecnologia e Inova\c{c}\~{a}o (Brazil) and Ministerio de Ciencia, Tecnolog\'{i}a e 
Innovaci\'{o}n Productiva (Argentina). 

This research has made use of the NASA/IPAC Extragalactic Database (NED) 
which is operated by the Jet Propulsion Laboratory, California Institute of 
Technology, under contract with the National Aeronautics and Space Administration.
We acknowledge the usage of the HyperLeda database (http://leda.univ-lyon1.fr).
{\sc chianti} is a collaborative project involving George Mason University, the 
University of Michigan (USA) and the University of Cambridge (UK). 



{\it Facilities:} \facility{Gemini-South (GMOS)}.

\end{document}